%% file: HIG-14-034_temp.tex
\pdfoutput=1

\documentclass[11pt,twoside,a4paper,cmspaper,final,collab]{cms-tdr}
\def\svnVersion{340276}\def\cmsCernNoTag{CERN-PH-EP/2015-211}\def\cmsCernDate{\today}\def\cmsMessage{Published in Physics Letters B as \href{http://dx.doi.org/10.1016/j.physletb.2016.01.056}{\doi{10.1016/j.physletb.2016.01.056.}}}

\begin{document}\cmsNoteHeader{HIG-14-034}

\hyphenation{had-ron-i-za-tion}
\hyphenation{cal-or-i-me-ter}
\hyphenation{de-vices}
\RCS$Revision: 331348 $
\RCS$HeadURL: svn+ssh://svn.cern.ch/reps/tdr2/papers/HIG-14-034/trunk/HIG-14-034.tex $
\RCS$Id: HIG-14-034.tex 331348 2016-03-08 17:01:40Z giuseppe $
\providecommand{\MT}{\ensuremath{m_\mathrm{T}}\xspace}
\newlength\cmsFigWidth
\ifthenelse{\boolean{cms@external}}{\setlength\cmsFigWidth{0.95\columnwidth}}{\setlength\cmsFigWidth{0.7\textwidth}}
\ifthenelse{\boolean{cms@external}}{\providecommand{\cmsLeft}{top\xspace}}{\providecommand{\cmsLeft}{left\xspace}}
\ifthenelse{\boolean{cms@external}}{\providecommand{\cmsRight}{bottom\xspace}}{\providecommand{\cmsRight}{right\xspace}}
\ifthenelse{\boolean{cms@external}}{\providecommand{\cmsTopLeft}{top\xspace}}{\providecommand{\cmsTopLeft}{top left\xspace}}
\ifthenelse{\boolean{cms@external}}{\providecommand{\cmsTopRight}{middle\xspace}}{\providecommand{\cmsTopRight}{top right\xspace}}
\newlength\cmsFigWidthThree
\ifthenelse{\boolean{cms@external}}{\setlength\cmsFigWidthThree{0.8\columnwidth}}{\setlength\cmsFigWidthThree{0.49\textwidth}}

\newcommand{\vecMET}{\ensuremath{\vec{E}_\mathrm{T}^\text{miss}}\xspace}
\providecommand{\tauh}{\ensuremath{\Pgt_\Ph}\xspace}
\providecommand{\CLs}{\ensuremath{\mathrm{CL}_\mathrm{s}}\xspace}

\newcommand{\PA}{\ensuremath{\cmsSymbolFace{A}}\xspace}
\providecommand{\Ph}{\ensuremath{\cmsSymbolFace{h}}\xspace}
\providecommand{\PZ}{\ensuremath{\cmsSymbolFace{Z}}\xspace}

\ifthenelse{\boolean{cms@external}}
{\providecommand{\suppMaterial}{the supplemental material
  [URL will be inserted by publisher]}}
{\providecommand{\suppMaterial}{Appendix~\ref{sec:kinfit}}}
\providecommand{\refreco}{Sec.~\ref{sec:reco}\xspace}

\cmsNoteHeader{HIG-14-034}
\title{Searches for a heavy scalar boson $\PH$ decaying to a pair of 125\GeV Higgs bosons
$\Ph\Ph$ or for a heavy pseudoscalar boson $\PA$ decaying to $\PZ\Ph$, in the final states with
$\Ph\to\Pgt \Pgt$}

\date{\today}

\abstract{
 A search for a heavy scalar boson $\PH$ decaying into a pair of lighter standard-model-like 125\GeV Higgs bosons $\Ph$ and a search for a heavy pseudoscalar boson $\PA$  decaying into a $\PZ$ and an $\Ph$ boson are presented. The searches are performed on a dataset corresponding to an integrated luminosity of 19.7\fbinv of pp collision data at a centre-of-mass energy of 8\TeV, collected by CMS
in 2012. A final state consisting of two $\tau$ leptons and two b jets is used to search for the $\PH\to \Ph\Ph$ decay. A final state
consisting of two $\tau$ leptons from the $\Ph$ boson decay, and two additional leptons from the $\PZ$ boson decay, is used to search for the decay $\PA\to\Z\Ph$.
The results are interpreted in the context of two-Higgs-doublet models.
No excess is found above the standard model expectation and upper limits are set on the heavy boson production cross sections in the mass ranges $260<m_{\PH}<350$\GeV and $220<m_{\PA}<350$\GeV.
 }

\hypersetup{%
pdfauthor={CMS Collaboration},%
pdftitle={Searches for a heavy scalar boson H decaying to a pair of 125 GeV Higgs bosons
hh or for a heavy pseudoscalar boson A decaying to Zh, in the final states with
h to tautau},%
pdfsubject={CMS},%
pdfkeywords={CMS, physics, MSSM, tau, higgs}}

\maketitle

\section{Introduction}
\label{sec:introduction}

The discovery of additional Higgs bosons at the LHC would provide
direct evidence of physics beyond the standard model (SM).
There are several types of models that require two Higgs doublets~\cite{Glashow:1976nt,Gunion:1989we,Branco:2011iw}. For example the minimal supersymmetric extension of the SM
(MSSM) requires the introduction of an additional Higgs doublet, where one Higgs doublet couples to up-type quarks and the other to down-type quarks~\cite{Fayet:1974pd,Fayet:1976et,Fayet:1977yc,Dimopoulos:1981zb,Sakai:1981gr,Inoue:1982ej,Inoue:1982pi,Inoue:1983pp}.
This leads to the prediction of five Higgs particles: one light and one heavy CP-even Higgs boson, \Ph and $\PH$,
one CP-odd Higgs boson
$\PA$, and two charged Higgs bosons $\PH^\pm$ ~\cite{Gunion:1989we,Djouadi:2005gj}. The masses and couplings of these bosons are interrelated and, at tree level, can be described by two parameters, which are often chosen to be the mass
of the pseudoscalar boson  $m_{\PA}$ and the ratio of the vacuum expectation values of the neutral components
of the two Higgs doublets $\tan\beta$. However, radiative corrections \cite{Okada:1990vk, Ellis:1990nz, Haber:1990aw, Carena:1995bx, Carena:1999py} introduce dependencies on other parameters namely
the mass of the top quark $m_{\cPqt}$, the scale of the soft supersymmetry breaking masses $M_{\mathrm{SUSY}}$,
the higgsino mass parameter $\mu$, the wino mass parameter $M_{2}$,
 the third-generation trilinear couplings, $A_{\cPqt}$, $A_{\cPqb}$, and $A_{\tau}$,
the mass of the gluino $m_{\PSg}$, and the third-generation slepton mass parameter $M_{\tilde{\ell}_{3}}$.

Direct searches for the neutral MSSM Higgs bosons have been performed by the CMS and ATLAS
Collaborations~\cite{Khachatryan:2014wca,Khachatryan:2015tra,Aad:2014vgg} using the benchmark scenarios proposed
in Ref.~\cite{Carena:2013qia}. In these scenarios the parameters involved in the radiative corrections for the Higgs
boson masses and couplings have been fixed, and only the two parameters $m_{\PA}$ and $\tan\beta$ remain
free. The value of $M_{\mathrm{SUSY}}$ was fixed at around 1\TeV, which produces a
lightest CP-even Higgs boson with a mass $m_{\Ph}$ lower than the observed Higgs boson mass of
$125.09 \pm 0.21\stat\pm 0.11\syst\GeV$~\cite{Aad:2015zhl}, for values of
$\tan\beta\lesssim6$.

If, however, $M_{\mathrm{SUSY}}$ is much larger than 1\TeV, as suggested by the non-observations of SUSY partner particles at the LHC so far,
low values of $\tan\beta$ can produce an $\Ph$ boson with $m_{\Ph}\simeq 125\GeV$~\cite{Djouadi:2013vqa,Carena:2014nza}. The interpretation of the Higgs boson measurements in the framework of the recently
developed MSSM benchmark scenarios ~\cite{Djouadi:2013uqa,Djouadi:2015jea,Carena:2014nza,LHCHXSWGnote}
suggests that the mass of the CP-odd Higgs boson, $m_{\PA}$, can be smaller than $2m_{\cPqt}$.
In the mass region below $2 m_{\cPqt}$ and at low values of $\tan\beta$, the decay mode of the
heavy scalar $\PH\to\Ph\Ph$ and that of the pseudoscalar $\PA\to\PZ\Ph$ can have sizeable branching fractions.

This encourages a programme of searches in the so-called ``low $\tan\beta$" channels~\cite{Djouadi:2013vqa,Arbey:2013jla}:

\begin{itemize}
\item{for $220\GeV<m_{\PA} < 2 m_{\cPqt}$:
      $\PA \to \PZ\Ph$;}
\item{for $260\GeV<m_{\PA} < 2 m_{\cPqt}$:
      $\PH \to \Ph\Ph$};
\item{for $m_{\PA} > 2 m_{\cPqt}$: $\PA/\PH \to \cPqt\cPaqt$.}
\end{itemize}

The decay modes $\PH\to\Ph\Ph$ and $\PA\to\PZ\Ph$, studied in this paper, are also present in other types of two-Higgs-doublet models (2HDM)~\cite{Gunion:1989we,Branco:2011iw}.
There are different types of 2HDM with those most similar to the MSSM (
i.e. where up-type fermions couple to one doublet and down-type fermions to the other) being ``Type II" 2HDM.
The discovery of a Higgs boson at the LHC~\cite{Aad:2012tfa,Chatrchyan:2012ufa,Chatrchyan:2013lba} with a mass around $125\GeV$ pushes the
2HDM parameter space towards either the alignment or decoupling limits ~\cite{Carena:2014nza}. In these limits
the properties of $\Ph$ are SM-like.

In the alignment limit of 2HDM when $\cos(\beta-\alpha)\ll 1$ (where $\alpha$ is the mixing angle
between the two neutral scalar fields), the $\PH\Ph\Ph$ and $\PA\PZ\Ph$ couplings
vanish at Born level~\cite{Asner:2013psa}. However, in the MSSM, the  $\PH\Ph\Ph$ and $\PA\PZ\Ph$ couplings do not vanish, even in the alignment limit,
because of the large radiative corrections that arise in the model.
In the decoupling limit of 2HDM the
scalar Higgs boson $\PH$ has a very large mass and the decay $\PH \to \cPqt\cPaqt$ dominates~\cite{Asner:2013psa}.

This paper reports the results of searches for the decays
$\PH\to\Ph\Ph\to\cPqb\cPqb\Pgt\Pgt$
and
$\PA\to\PZ\Ph\to{\ell\ell}\Pgt\Pgt$ (where $\ell\ell$ denotes $\Pgm\Pgm$ or
$\Pe\Pe$). The choice of $\Pgt$ pair final state was driven by its quite clean
signature and by the most recent results, which gave stronger
evidence of the 125 Higgs boson coupling to the fermions~\cite{Chatrchyan:2014vua}. This analysis exploits similar techniques as used for the search
for the SM Higgs boson at 125\GeV~\cite{Chatrchyan:2014nva} and
several different $\Pgt\Pgt$ signatures are studied. For the channel $\PH\to\Ph\Ph\to\cPqb\cPqb\Pgt\Pgt$, the $\Pgm\tauh$, $\Pe\tauh$, and $\tauh\tauh$ final states are used, where $\tauh$ denotes the visible products of a hadronically decaying $\Pgt$,  whereas for
the channel $\PA\to\PZ\Ph\to{\ell\ell}\Pgt\Pgt$,
the $\Pgm\tauh$, $\Pe\tauh$, $\tauh\tauh$, and $\Pe\Pgm$ final states are selected.

Searches for the decays $\PH \to \Ph\Ph$,
and $\PA \to \PZ\Ph$ have already been performed by the
ATLAS~\cite{Aad:2014yja,Aad:2015wra,Aad:2015uka,Aad:2015xja} and CMS
Collaborations~\cite{Khachatryan:2014jya,Khachatryan:2015wwa,Khachatryan:2015yjb}
in di-photon, multilepton and $\cPqb\cPqb$ final states.

This analysis has the power to bring important results in the low $\tan\beta$ region for the m$_{\PA}$ range, which has been previously discussed and where these processes have an enhanced sensitivity~\cite{Djouadi:2013vqa}. This region has not yet been excluded by the direct or indirect searches for a heavy scalar or pseudoscalar Higgs boson, that have been mentioned above, therefore the described decay modes look to be quite promising.

For simplicity of the paper, we are neither indicating the charge of the leptons nor the particle-antiparticle nature of quarks.

\section{The CMS detector, simulation and data samples}
\label{sec:cmsdet}

A detailed description of the CMS detector can be found in Ref.~\cite{Chatrchyan:2008aa}.
The central feature of the CMS apparatus is a superconducting solenoid of 6\unit{m} internal diameter providing a field of
3.8\unit{T}. Within the field volume are a silicon pixel and strip tracker, a crystal electromagnetic calorimeter (ECAL), and a brass/scintillator
hadron calorimeter. Muons are measured in gas-ionisation detectors embedded in the steel return yoke of the magnet.

The CMS coordinate system has the origin centered at the nominal collision point and is oriented such that the $x$-axis points to the center
of the LHC ring, the $y$-axis points vertically upward and the $z$-axis
is in the direction of the beam. The azimuthal angle $\phi$ is measured from the $x$-axis in the
$xy$ plane and the radial coordinate in this plane is denoted by $r$. The polar angle $\theta$ is
defined in the $rz$ plane and the pseudorapidity is $\eta= -\ln [ \tan(\theta/2)]$~\cite{Chatrchyan:2008aa}. The momentum component transverse to the beam direction, denoted by \PT, is computed from the $x-$ and $y-$components.

The first level (L1) of the CMS trigger system, composed of custom hardware processors, uses information from the calorimeters and muon detectors
to select the most interesting events in a fixed time interval of less than 4\mus. The high-level Trigger processor farm decreases
the L1 accept rate from around 100\unit{kHz} to less than 1\unit{kHz} before data storage.

The data used for this search were recorded with the CMS detector
 in proton-proton collisions at the CERN LHC
and correspond to an integrated luminosity of 19.7\fbinv at a centre-of-mass energy of $\sqrt{s}=8$\TeV.
The $\PH\to\Ph\Ph$ signals are modelled with the \PYTHIA\,6.4.26~\cite{Sjostrand:2006za} event generator while the $\PA\to\PZ\Ph$ signals were modelled with
  \MADGRAPH\,5.1~\cite{Alwall:2011uj}.
When modelling background processes, the \MADGRAPH\,5.1 generator is used for $\cPZ$+jets, $\PW$+jets, $\cPqt\cPaqt$, and diboson production, and {\POWHEG}\,1.0 \cite{Nason:2004rx,Frixione:2007vw,Alioli:2009je,Alioli:2010xd} for single top quark production.
The \POWHEG and \MADGRAPH generators are interfaced with \PYTHIA for parton showering and fragmentation using the Z2* tune \cite{Chatrchyan:2013ala}. All generators are interfaced with  \TAUOLA~\cite{Davidson:2010rw} for the simulation of the $\Pgt$ decays.  All generated events are processed through a detailed simulation of the CMS detector based on {\GEANTfour}~\cite{Agostinelli:2002hh} and are reconstructed with the same algorithms as the data.
Parton distribution functions (PDFs) CT10 \cite{Lai:2010vv} or CTEQ6L1 \cite{Pumplin:2002vw} for the proton are used, depending on the generator in question, together with MSTW2008 \cite{Martin:2009iq} according to PDF4LHC prescriptions \cite{Botje:2011sn}.

\section{Event reconstruction}
\label{sec:reco}

During the 2012 LHC run there were an average of 21 proton-proton interactions per bunch crossing.
The collision vertex that maximizes the sum of the squares of momenta components perpendicular to the beamline
(transverse momenta) of all tracks associated with it, $\sum\PT^2$, is taken to be the vertex of the primary hard interaction.
 The other vertices are categorised as pileup vertices.

A particle-flow algorithm~\cite{CMS-PAS-PFT-09-001,CMS-PAS-PFT-10-001} is used to
reconstruct individual particles, \ie muons, electrons, photons, charged hadrons and neutral hadrons, using
information from all CMS subdetectors. Composite objects
such as jets, hadronically decaying $\Pgt$ leptons, and missing transverse energy are then constructed using the lists of individual particles.

Muons are reconstructed by performing a simultaneous global track fit to hits in the silicon tracker and the muon system~\cite{Chatrchyan:2012xi}.
Electrons are reconstructed from clusters of ECAL energy deposits matched
to hits in the silicon tracker~\cite{Khachatryan:2015hwa}.
Muons and electrons assumed to originate from $\PW$ or $\cPZ$ boson decays are required to be spatially isolated
from other particles ~\cite{Khachatryan:2015hwa, Chatrchyan:2013sba}.
The presence of charged and neutral particles from pileup vertices is taken into account in the isolation requirement of both muons and electrons.
Muon and electron identification and isolation efficiencies are measured via the tag-and-probe technique~\cite{Khachatryan:2010xn} using inclusive
samples of $\PZ\to\ell\ell$ events from data and simulation. Correction factors are applied to account for differences between data and simulation.

Jets are reconstructed from all particles using the anti-$\kt$ jet clustering algorithm
implemented in \textsc{fastjet}~\cite{Cacciari:fastjet1, Cacciari:fastjet2} with a
distance parameter of 0.5.
The contribution to the jet energy from particles originating from pileup vertices is
removed following a procedure based on the effective jet area described in Ref. ~\cite{Cacciari:2008:puJetArea}.
Furthermore, jet energy corrections are applied as a function of jet \PT and $\eta$ correcting jet energies to the generator level response of the jet, on average.
Jets originating from pileup interactions are removed by a multivariate pileup jet identification algorithm~\cite{CMS-PAS-JME-13-005}.

The missing transverse momentum vector $\ptvecmiss$ is defined as the negative vector sum
of the transverse momenta of all reconstructed particles in the volume of the detector (electrons, muons, photons, and hadrons). Its magnitude is referred to as \MET.
The \ETmiss reconstruction is improved by taking into account the jet energy scale corrections and the $\phi$ modulation, due to collisions not being at the nominal centre of CMS~\cite{Khachatryan:2014gga}. A multivariate regression correction of \ETmiss, where the contributing particles are separated into those coming from the primary vertex and those that are not, mitigates the effect of pileup~\cite{Khachatryan:2014gga}.

Jets from the hadronisation of b-quarks (b jets) are identified with the
combined secondary vertex (CSV) b tagging algorithm~\cite{Chatrchyan:2012jua}, which exploits the
information on the decay vertices of long-lived mesons and the transverse impact parameter
measurements of charged particles.
This information is combined in a likelihood discriminant. The medium value of the CSV discriminator, corresponding to a b jet misidentification probability of 1\%, has been used in this analysis.

Hadronically decaying $\Pgt$ leptons are reconstructed using the hadron-plus-strips algo\-rithm \cite{Khachatryan:2015dfa}, which considers candidates with one charged pion and up to two neutral pions, or three charged pions.
 The neutral pions are reconstructed as ``strips'' of electromagnetic
particles taking into account possible broadening of calorimeter energy
depositions in the $\phi$ direction from photon conversions. The \tauh candidates that are also compatible with muons or electrons are rejected.
Jets originating from the hadronization of quarks and gluons are suppressed by requiring the \tauh candidate to be
isolated. The contribution of charged and neutral particles from pileup interactions is removed when computing the isolation.
\section{Event selection}
\label{sec:eventSelection}

The events are selected with a combination of electron, muon and $\Pgt$ trigger objects ~\cite{Khachatryan:2015hwa,Chatrchyan:2013sba,Chatrchyan:2011nv,Chatrchyan:2014nva}.
The identification criteria of these objects were progressively tightened and their transverse momentum thresholds  raised as the
LHC instantaneous luminosity increased over the data taking period. A tag-and-probe method was used to measure the efficiencies of these triggers in
data and simulation, and correction factors are applied to the simulation.

Electrons, muons, and $\tauh$ are selected using the criteria defined in the CMS search for the SM Higgs boson at 125\GeV~\cite{Chatrchyan:2014nva}.
Specific requirements for the selection of the $\PH\to\Ph\Ph\to\cPqb\cPqb\Pgt\Pgt$ and the $\PA\to\PZ\Ph\to{\ell\ell}\Pgt\Pgt$ channels are described below.

\subsection{Event selection of \texorpdfstring{$\PH\to\Ph\Ph\to\cPqb\cPqb\Pgt\Pgt$}{H to  hh to bbtt}}

In the $\PH\to\Ph\Ph\to\cPqb\cPqb\Pgt\Pgt$ channel, the three most sensitive final states are analysed, distinguished by the decay mode of the two $\Pgt$ leptons originating from the h boson ($\Pgm\tauh$, $\Pe\tauh$ and $\tauh\tauh$).

 In the $\Pgm\tauh$ and $\Pe\tauh$ final states, events are selected with a muon with $\pt>20\GeV$ and $\abs{\eta} < 2.1$ or an electron of $\pt>24\GeV$ and $\abs{\eta} < 2.1$, and an oppositely charged $\tauh$ of $\pt > 20\GeV$ and $\abs{\eta} < 2.3$.
To reduce the $\cPZ\to\Pgm\Pgm, \Pe\Pe$ contamination, events with two muons or electrons of $\pt >15\GeV$, of opposite charges, and passing loose isolation criteria are rejected.

\begin{figure}[htb]
\centering
\includegraphics[width=\cmsFigWidth]{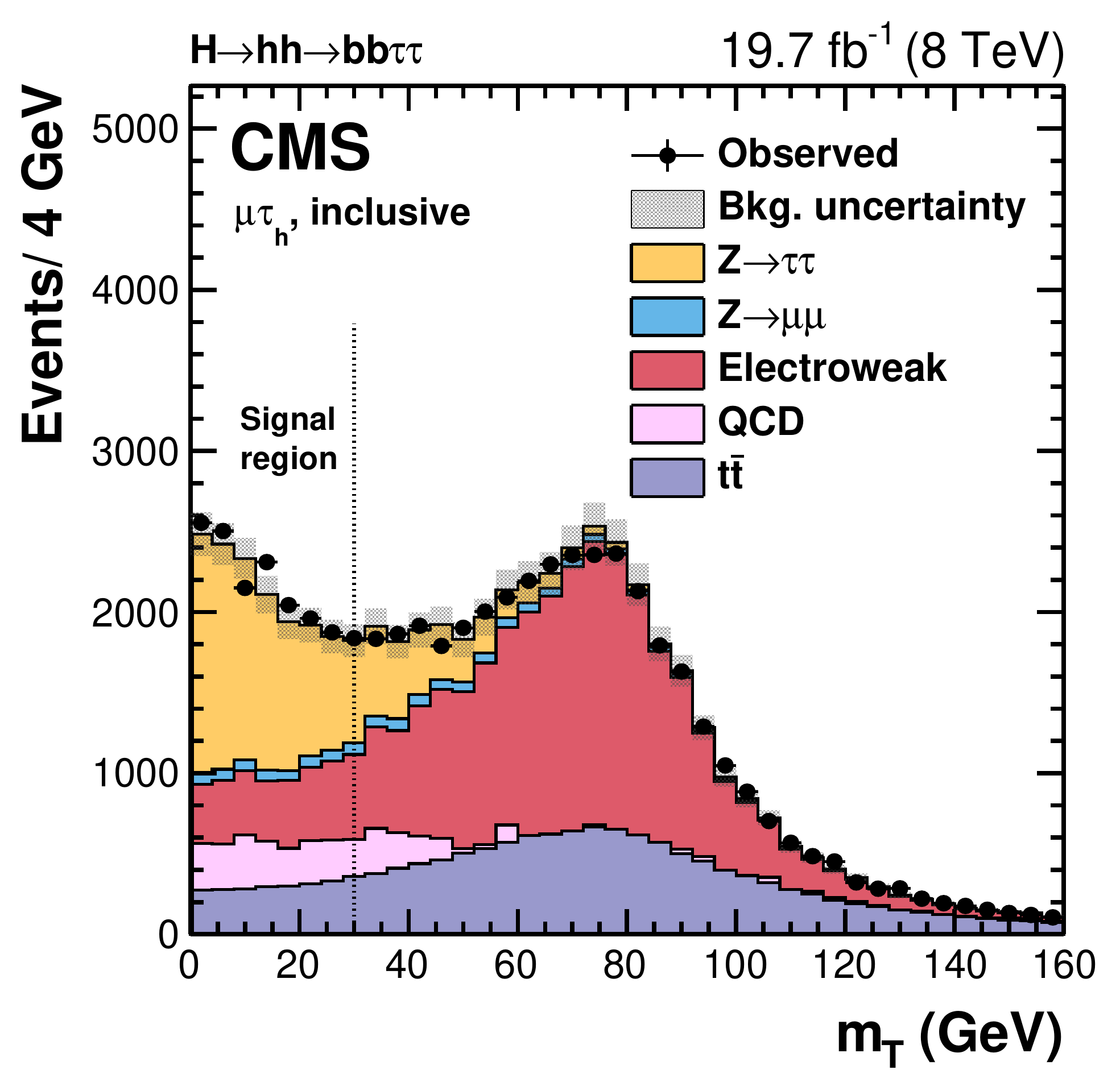}
\caption{Distribution of $\MT$ for events in the $\Pgm\tauh$ final state, containing at least two additional jets.
The $\PW$+jets background is included in the ``electroweak" category. Multijet events are indicated as QCD.
 The $\PH\to\Ph\Ph\to\cPqb\cPqb\tau\tau$ selection requires $\MT < 30\GeV$ for the $\Pgm\tauh$ and $\Pe\tauh$ final states.}
\label{fig:mt_mutau}
\end{figure}

 In the $\Pgm\tauh$ and $\Pe\tauh$ final states, the transverse mass of the muon or electron and $\vec{p}_{\mathrm{T}}^{\text{miss}}$ \\
\begin{equation}
\MT = \sqrt{2 \pt \MET (1-\cos\Delta\phi)},
\label{eq:mt}
\end{equation}
where $\pt$ is the lepton transverse momentum and $\Delta\phi$ is the difference in the azimuthal angle between the lepton momentum and $\vec{p}_{\mathrm{T}}^{\text{miss}}$, is required to be less than 30\GeV to reject events coming from $\PW$+jets and $\cPqt\cPaqt$ backgrounds.
The $m_T$ distribution for the $\Pgm\tauh$ final state is shown in Fig.~\ref{fig:mt_mutau}.

In the $\tauh\tauh$ final state, events with two oppositely charged hadronically decaying $\Pgt$ leptons with $\pt > 45\GeV$ and $\abs{\eta} < 2.1$ are selected.

In addition to the $\Pgt\Pgt$ selection, each selected event must contain at least two jets with $\pt>20\GeV$ and $|\eta|<2.4$.
These $\pt$ and $\eta$ requirements
are necessary to select jets that have a well defined value of the CSV discriminator (Section~\ref{sec:reco}), which is important for
categorising signal-like events with two b jet candidates coming from the
125\GeV Higgs boson decaying to $\cPqb\cPqb$.

Simulation studies show that the majority of signal events will
have at least one jet passing the medium working point of the CSV discriminator.
The jets are ordered by CSV discriminator value, such that the leading and
subleading jets are defined as those with the two highest CSV values. Then the
events are separated into categories, defined as:
\begin{itemize}
\item{2jet--0tag}
when neither the leading nor subleading jets
passes the medium CSV working point. Only a small amount of signal is collected
in this category, which is background-dominated.
\item{2jet--1tag} when only the leading but not the subleading jet passes the medium CSV working point.
\item{2jet--2tag} when both the leading and subleading jets
pass the medium CSV working point.
\end{itemize}

The signal extraction is performed using the distribution of the reconstructed mass of the H boson candidate.

\subsection{Event selection of \texorpdfstring{$\PA\to\PZ\Ph\to{\ell\ell}\Pgt\Pgt$}{A to Zh to ell ell tau tau}}

In the $\PA\to\PZ\Ph\to{\ell\ell}\Pgt\Pgt$ channel eight final states are analysed.
These are categorised according to the decay mode of the Z boson and the decay mode of the $\Pgt$ leptons
originating from the $\Ph$ boson.

The $\PZ$ boson is reconstructed from two same-flavour, isolated, and oppositely charged electrons or muons. In the $\PZ\to\Pgm\Pgm$ ($\Pe\Pe$) final state the muons (electrons) are required to have $|\eta|<2.4$ (2.5) with $\pt>20$\GeV  for the leading lepton and $\pt>10$\GeV for the subleading lepton. The invariant mass of the two leptons is required to be between 60\GeV and 120\GeV. When more than one pair of leptons satisfy these criteria, the pair with an invariant mass closest to the Z boson mass is selected.

After the $\PZ$ candidate has been chosen, the $\Ph\to\Pgt\Pgt$ decay is selected by combining the decay products of the two $\Pgt$ leptons in the four final states $\Pgm\tauh, \Pe\tauh, \tauh\tauh, \Pe\Pgm$.
The combination of the large contribution from the irreducible $\PZ\PZ$ background and of the small branching fractions of leptonic tau decays
makes the $\Pgm\Pgm$ and $\Pe\Pe$ final states less sensitive to the signal, and therefore they are not used in the analysis.
Depending on the final state, a muon with $\pt>10\GeV$ and $\abs{\eta}<2.4$, or an electron of $\pt>10\GeV$ and $\abs{\eta}<2.5$,
or a $\tauh$ of $\pt>21\GeV$ and $\abs{\eta}<2.3$ are combined to form an oppositely charged pair. Events with additional light leptons satisfying these requirements are rejected.

\begin{figure}[htb]
\centering
\includegraphics[width=\cmsFigWidth]{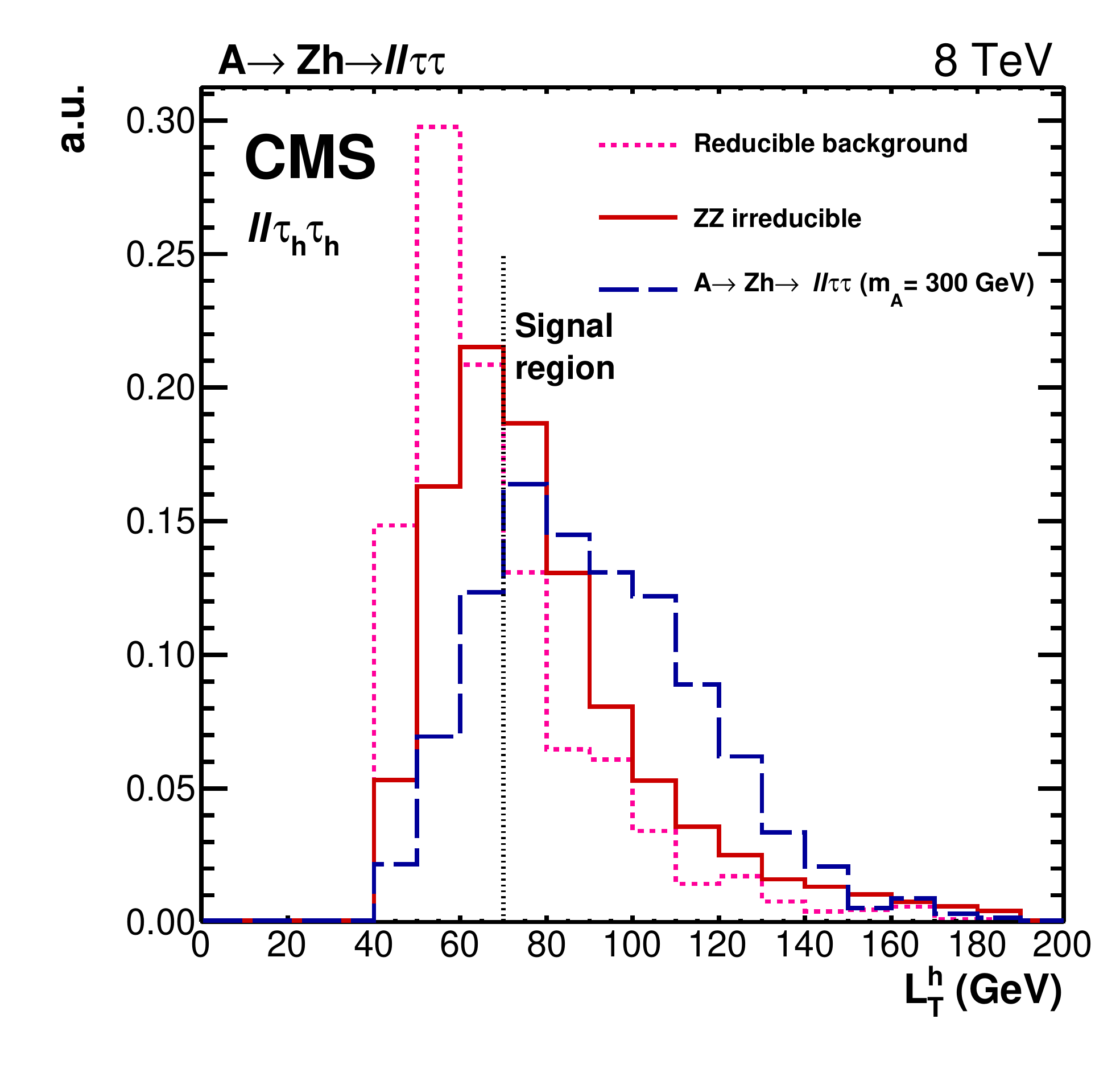}
\caption{Distribution of the variable $L_\mathrm{T}^{\Ph}$ for events in the ${\ell\ell}\tauh\tauh$ final state. The reducible background is estimated from data, instead the ZZ irreducible background from simulation.
}
\label{fig:lt}
\end{figure}

A requirement on $L_\mathrm{T}^{\Ph}$, which is the scalar sum of the visible transverse momenta of the two $\Pgt$ candidates originating from the $\Ph$ boson, is applied to lower the reducible background from misidentified leptons as well as the irreducible background from $\PZ\PZ$ production.
The thresholds of this requirement depend on the final state and have been chosen
in such a way as to optimise the sensitivity of the analysis to the presence of an $\PA\to\PZ\Ph$ signal for $\PA$ masses between 220 and 350\GeV. The distribution of $L_\mathrm{T}^{\Ph}$ for events in the ${\ell\ell}\tauh\tauh$ final state can be seen in Fig.~\ref{fig:lt}.

In order to reduce the $\cPqt\cPaqt$ background, events containing a jet  with $\pt>20\GeV$, $\abs{\eta}<2.4$ and passing the medium working point of the CSV b tagging discriminator are removed.

The four final objects are further required to be separated from each other by \DR = $\sqrt{\smash[b]{(\Delta \eta)^2 + (\Delta \varphi)^2}}$ larger than 0.5 (where phi is in radians), and to come from the same primary vertex.

In this channel the signal extraction is performed using the distribution of the reconstructed mass of the A boson candidate.

\section{Background estimation}
\label{sec:background}

\subsection{Background estimation for \texorpdfstring{$\PH\to\Ph\Ph\to\cPqb\cPqb\Pgt\Pgt$}{H to hh to bb tau tau}}
\label{sec:background_hh}

The backgrounds to the $\PH\to\Ph\Ph\to\cPqb\cPqb\Pgt\Pgt$ final state consist predominantly  of $\cPqt\cPaqt$ events, followed by
$\Z \to \Pgt\Pgt$+jets events, $\PW$+jets events, and QCD multijet events, with other small contributions from $\Z \to
\ell\ell$, diboson, and single top quark production.
The estimation of the shapes of the reconstructed H mass and of the yields of the major backgrounds is obtained from data wherever possible.

The $\cPZ\to\Pgt\Pgt$ process constitutes an irreducible background due to its final
state involving two $\Pgt$ leptons, which only differ from the $\Ph\to\Pgt\Pgt$ signal by
having an invariant mass closer to the mass of the $\PZ$ boson instead of the Higgs boson.
Requiring two jets in the event greatly reduces this background and the b tagging requirements reduce it even further.
Nevertheless, it still remains an important source of background events,
in particular in the 2jet--1tag and 2jet--0tag categories.
This background is estimated using a sample of $\cPZ\to\Pgm\Pgm$ events from data,
obtained by requiring two oppositely charged isolated muons, where the reconstructed muons are replaced  by the reconstructed  particles from simulated $\Pgt$ decays. A correction for a contamination from $\cPqt\cPaqt$ events is applied to the $\cPZ\to\Pgm\Pgm$ selection. This technique substantially reduces the systematic uncertainties
due to the jet energy scale and the missing transverse energy, as these quantities are modelled with
data.

For the $\cPqt\cPaqt$ background, both shape and normalisation are taken
from Monte Carlo simulation (MC), and the results are checked
against data in a control region where the presence of $\cPqt\cPaqt$
events is enhanced by requiring $\Pe\Pgm$ in the final state instead of a ditau, and at least one b tagged jet.

Another significant source of background is from QCD multijet events, which can
mimic the signal in various ways, e.g. where one or more jets are misidentified as $\tauh$.
In the $\Pgm\tauh$ and $\Pe\tauh$ channels, the shape of the QCD background is estimated using an observed sample of same-sign (SS)
$\Pgt\Pgt$ events. The yield is obtained by scaling the observed number of SS
events by the ratio of the opposite-sign (OS) to SS event yields obtained in a
QCD-enriched region with relaxed lepton isolation. In the $\tauh\tauh$ channel,
the shape is obtained from OS events with relaxed $\Pgt$ isolation. The yield is
obtained by scaling these events by the ratio of SS events with tighter and
relaxed $\Pgt$ isolation.

In the $\Pgm\tauh$ and $\Pe\tauh$ channels, $\PW$+jets events in which there is
a jet misidentified as a $\tauh$ are another sizeable source of background. The
$\PW$+jets shape is modelled using MC simulation and the yield is estimated using
a control region of events with large $m_T$ close to the $\PW$ mass. In the $\tauh\tauh$ channel this background has been found to be less relevant and its shape and yield are taken from MC simulation.

The contribution of Drell--Yan production of muon and electron pairs is estimated from simulation after rescaling the simulated yield to that measured from observed $\cPZ\to\Pgm\Pgm$ events. In the $\Pe\tauh$ channel, the $\cPZ\to\Pe\Pe$ simulation is further corrected using the $\Pe\to\tauh$ misidentification rate measured in data using a tag-and-probe technique~\cite{Khachatryan:2010xn} on $\cPZ\to\Pe\Pe$ events.

Finally the contributions of other minor backgrounds such as diboson and single
top quark events are estimated from simulation. Possible contributions from SM
Higgs boson production are estimated and found to have a negligible effect on
the final result.

\subsection{Background estimation for \texorpdfstring{$\PA\to\PZ\Ph\to\ell\ell\Pgt\Pgt$}{A to ZH to ell ell tau tau}}

The backgrounds to the $\PA\to\PZ\Ph$ channel can be divided into a reducible
component and an irreducible component which contribute in equal parts.

The predominant source of irreducible background is from $\PZ\PZ$ production that yields exactly the same final states as the expected signal. Other ``rare" sources of irreducible background are SM Higgs boson associated production with a $\PZ$ boson, $\cPqt\cPaqt\PZ$ production where the $\PZ$ boson decays into a muon or an electron pair and both top quarks decay leptonically (to~$\Pe$, $\Pgm$, or $\tauh$), and triboson events (WWZ, WZZ, ZZZ).
The contributions of all the irreducible backgrounds after the final selection are estimated from simulation.

The reducible backgrounds have at least one lepton in the final state that is due to a misidentified jet that passes the lepton identification. In $\ell\ell\tauh\tauh$ final states, the reducible background is essentially composed of Z+jets events with at least two jets, whereas in $\ell\ell\Pgm\tauh$ and $\ell\ell\Pe\tauh$ final states, the main contribution to the reducible background comes from WZ+jets with three light leptons. The contribution from these processes to the final selected events is estimated using control samples in data.

The probabilities for a jet that passes relaxed lepton selection criteria to pass the final identification and isolation criteria of electrons, muons, and $\Pgt$ leptons are measured in a signal-free region as a function of the transverse momentum of the object closest to the candidate, $f(\pt^{\text{fake}})$. In this region, events are required to pass all the final state selections, except that the reconstructed $\Pgt$ candidates are required to have the same sign and to pass relaxed identification and isolation criteria. This effectively eliminates any possible signal, while maintaining roughly the same proportion of reducible background events.

In order to use the misidentification probabilities $f(\pt^{\text{fake}})$, sidebands are defined for each channel, where, unlike the relaxed criterion, the final identification or isolation criterion is not satisfied for one or more of the final state lepton candidates. The number of reducible background events due to a lepton being misidentified in the final selection is estimated by applying the weight $f(\pt^{\text{fake}})/(1-f(\pt^{\text{fake}}))$ to the observed events with lepton candidates in the sideband that satisfy the relaxed but not the final identification or isolation criterion. Finally, the reducible background shape of the reconstructed $\PA$ mass is obtained from a SS signal--free region where the $\Pgt$ candidates have the same charge and relaxed isolation criteria. Possible contributions from SM Higgs boson production are estimated and found to have a negligible effect on the final result.

\section{Systematic uncertainties}
\label{sec:systematics}

The shape of the reconstructed mass of the $\PA$ and $\PH$ boson candidates, used for signal extraction, and the normalisation are sensitive to various systematic uncertainties.

The main contributions to the normalisation uncertainty that affect the signal and the simulated backgrounds include the uncertainty in the total integrated luminosity, which amounts to 2.6\%~\cite{CMS-PAS-LUM-13-001}, and the identification and trigger efficiencies of muons (2\%) and electrons (2\%). The $\tauh$ identification efficiency has a 6$\%$ uncertainty (8\% in the $\tauh\tauh$ channel), which is measured in
$\PZ/\gamma^{*} \to\Pgt\Pgt\to\Pgm\tauh$ events using a tag-and-probe technique. There is a 3\% uncertainty in the efficiency on the hadronic part of the $\mu\tauh$ and $\Pe\tauh$
triggers, and a 4.5\% uncertainty on each of the two $\tauh$ candidates
required by the $\tauh\tauh$ trigger. The b tagging efficiency has an
uncertainty of 2--7\%, and the mistag rate for light-flavour partons is accurate
to 10--20\% depending on $\eta$ and $\pt$~\cite{Chatrchyan:2012jua}.
The background normalisation uncertainties from the estimation methods discussed in Section~\ref{sec:background} are also considered. In the $\PH\to\Ph\Ph\to\cPqb\cPqb\Pgt\Pgt$ channel this uncertainties amount to 2--40\% depending on the event category and on the final state.
The uncertainties of reducible backgrounds to the  $\PA\to\PZ\Ph$ channel  are estimated by evaluating an individual uncertainty
for each lepton misidentification rate and  applying it to the background calculation. This amounts to 15--50\% depending on the final $\ell\ell\Pgt\Pgt$ state considered.
The main uncertainty in the estimation of the $\PZ\PZ$ background arises from the theoretical uncertainty in the $\PZ\PZ$ production cross section.

Uncertainties that contribute to variations in the shape of the mass spectrum include the jet energy scale, which varies with jet \pt and jet
$\eta$~\cite{Chatrchyan:2011ds}, and the $\Pgt$ lepton (3\%) energy scale~\cite{Chatrchyan:2014nva}.

Theoretical uncertainties on the cross section for signal derive from PDF and QCD scale uncertainties and depend on the choice of signal hypothesis. For model
independent results no choice of cross section is made and hence no theoretical uncertainties are considered. For the MSSM interpretation the uncertainties
depend on $m_{\PA}$ and $\tan\beta$ and amount to 2--3\% for PDF uncertainties and 5--9\% for scale uncertainties, evaluated as described in
\cite{LHCHXSWGnote} and using the PDF4LHC recommendations \cite{Botje:2011sn}. No theoretical uncertainties are considered in the 2HDM interpretation.
\section{Results and interpretation}
\label{sec:results}

The ditau ($m_{\Pgt\Pgt}$) mass is reconstructed using a dedicated algorithm
called \textsc{SVFit}~\cite{Bianchini:2014vza}, which combines the visible four-vectors of the $\Pgt$ lepton candidates as well as the \ETmiss and its experimental resolution in a maximum likelihood estimator.

For the $\PH\to\Ph\Ph\to\cPqb\cPqb\Pgt\Pgt$ process, the chosen
distribution for signal extraction is the four-body mass.
The decay products of the two $\Ph$ bosons need to fulfill stringent kinematic constraints, due to the small natural width of the $\Ph$. These constraints can be used in a kinematic fit in order to improve the event reconstruction and to better separate signal events from background.
The collinear approximation for the decay products of the $\Pgt$~leptons is assumed in the fit, since the $\Pgt$ leptons are highly boosted as they originate from an object that is heavy when compared to their own mass.
Furthermore, it is assumed that the reconstruction of the directions of all final state objects is accurate and the uncertainties can be neglected compared to the uncertainties on the energy reconstruction.
In the decay of the two $\Pgt$~leptons, at least two neutrinos are involved and there is no precise
measurement of the original $\Pgt$~lepton energies.
For this reason, the $\Pgt$~lepton energies are constrained from the balance of the fitted $\PH$ boson transverse momentum and the reconstructed transversal recoil determined from \MET reconstruction algorithms, as described in \refreco. The reconstructed mass obtained with the kinematic fit is denoted by $m_{\PH}^{\text{kinfit}}$  (see~\suppMaterial~for a detailed description).

\begin{figure}[htbp]
\centering
\includegraphics[width=\cmsFigWidthThree]{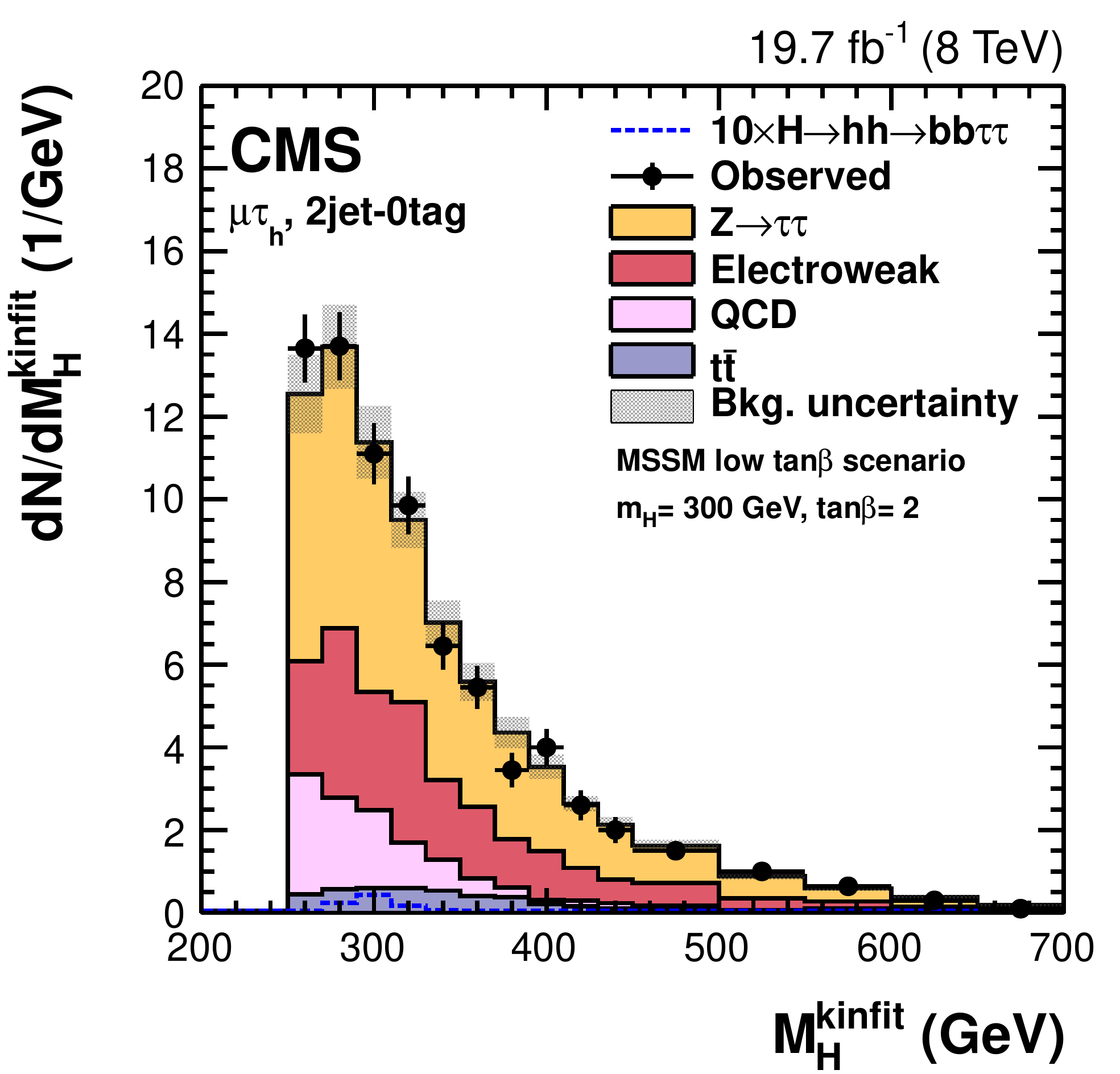}
\includegraphics[width=\cmsFigWidthThree]{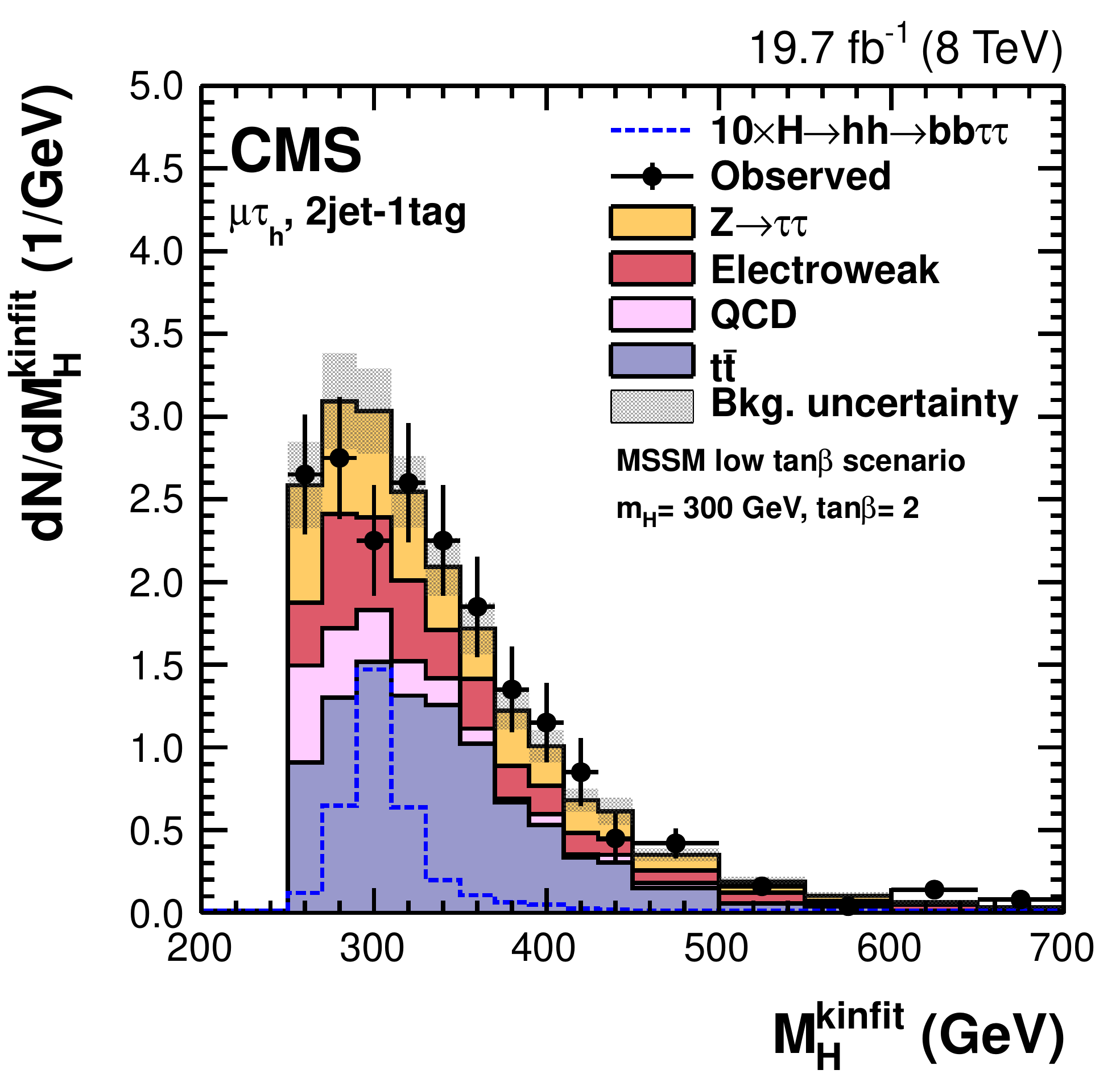}
\includegraphics[width=\cmsFigWidthThree]{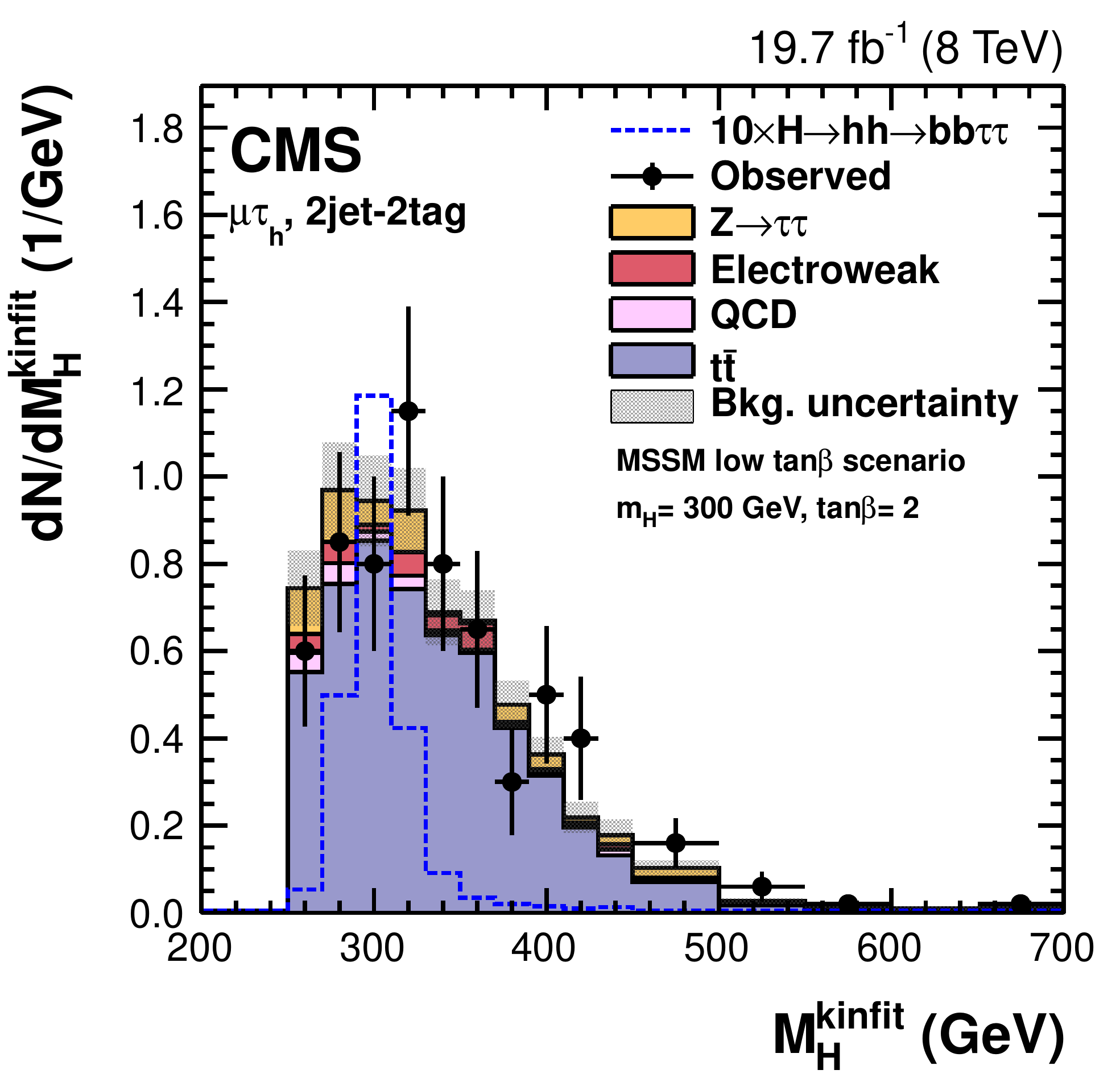}
\caption{
  Distributions of the reconstructed four-body mass with the
  kinematic fit after applying mass selections on $m_{\Pgt\Pgt}$ and $m_{\cPqb\cPqb}$
  in the $\Pgm\tauh$
  channel. The plots are shown for events in the 2jet--0tag (\cmsTopLeft), 2jet--1tag
  (\cmsTopRight), and 2jet--2tag (bottom) categories. The expected signal scaled by a
factor 10 is shown superimposed as an open dashed histogram for $\tan\beta = 2$
and $m_{\PH}= 300$\GeV in the low $\tan\beta$ scenario of the MSSM.
  Expected background contributions are shown for the values of nuisance parameters (systematic uncertainties)
  obtained after fitting the signal plus background hypothesis to the data.
}
\label{fig:resultsKinFitMassCutsMuTau}
\end{figure}

\begin{figure}[htbp]
\centering
\includegraphics[width=\cmsFigWidthThree]{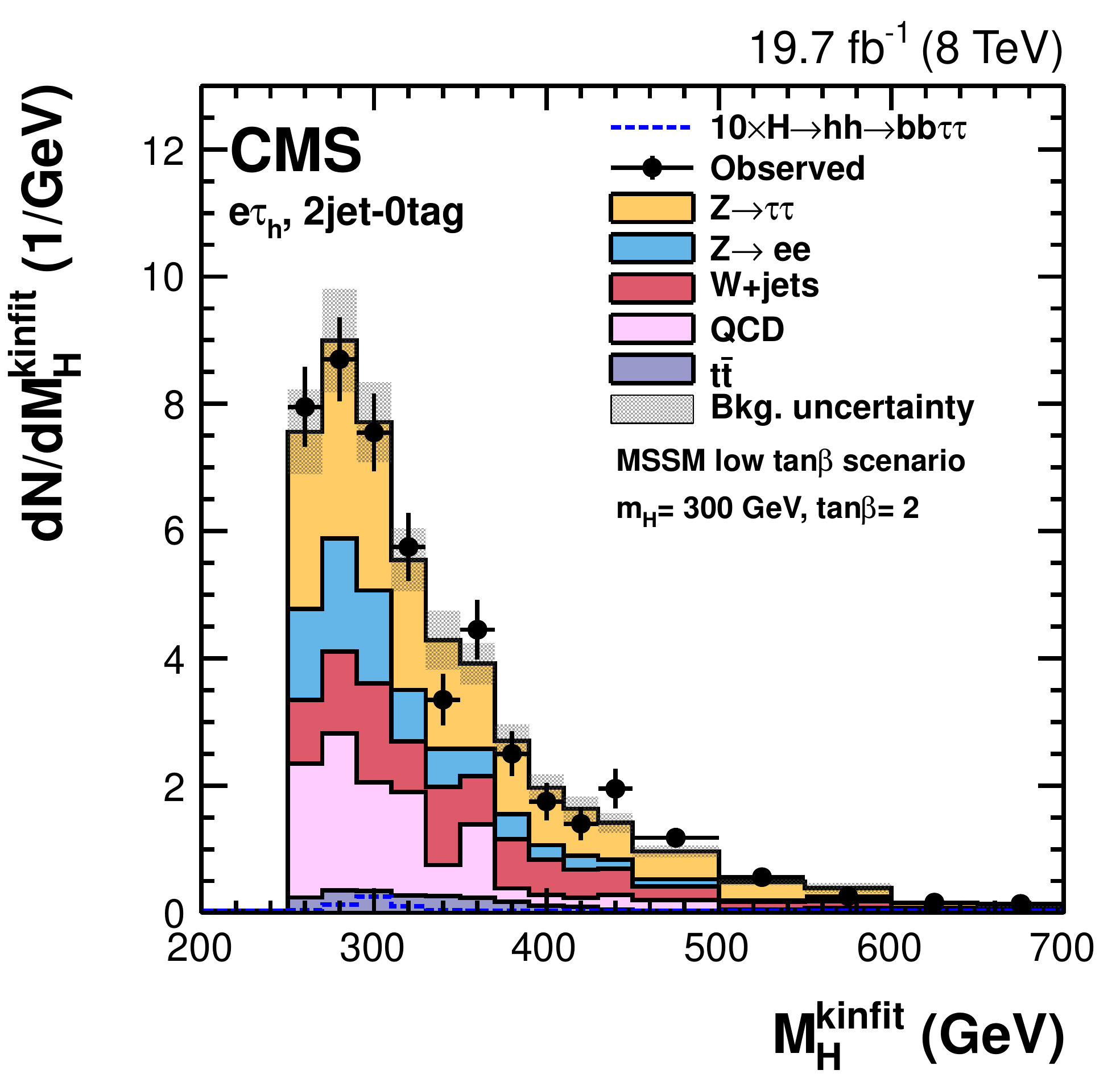}
\includegraphics[width=\cmsFigWidthThree]{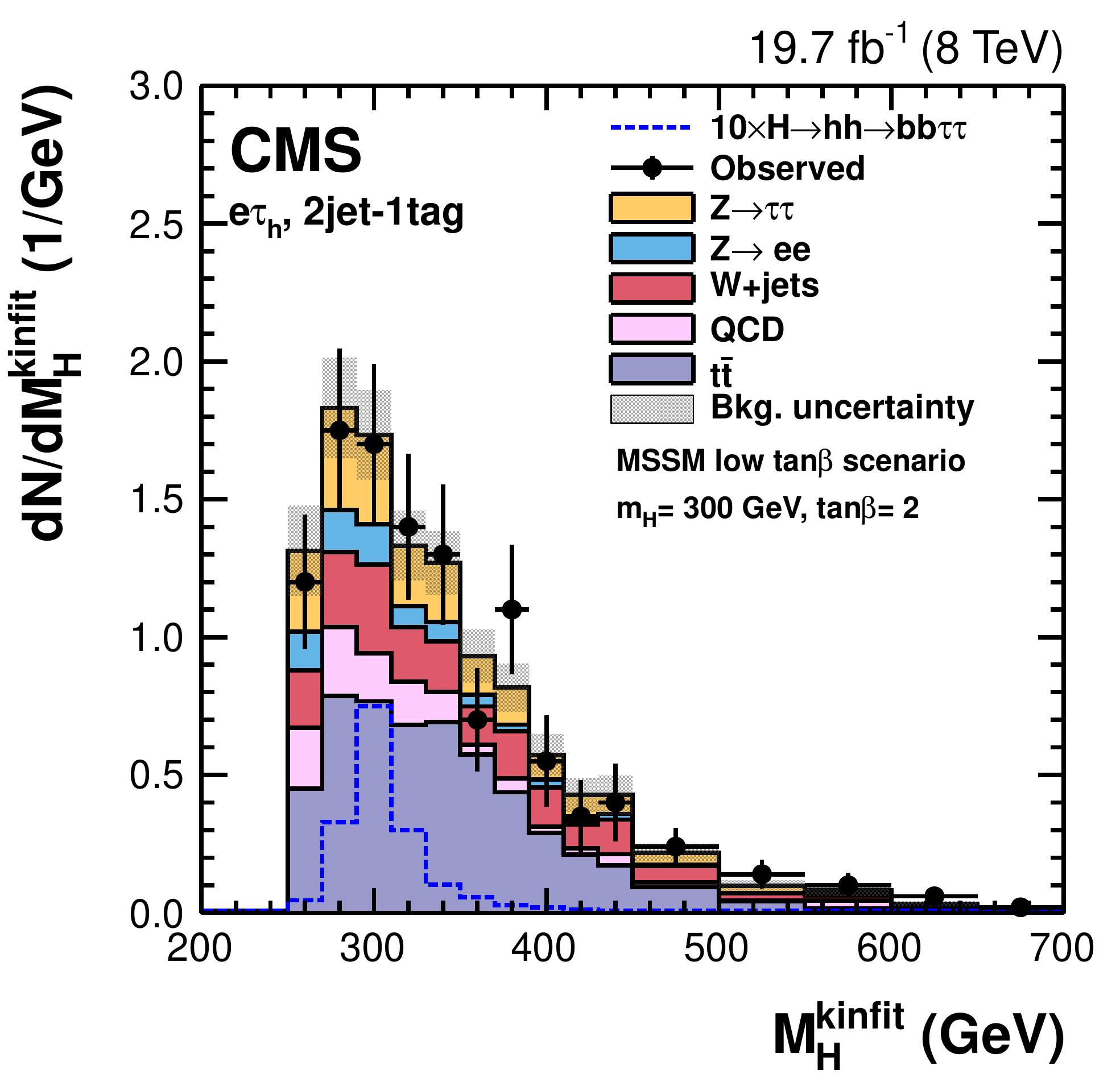}
\includegraphics[width=\cmsFigWidthThree]{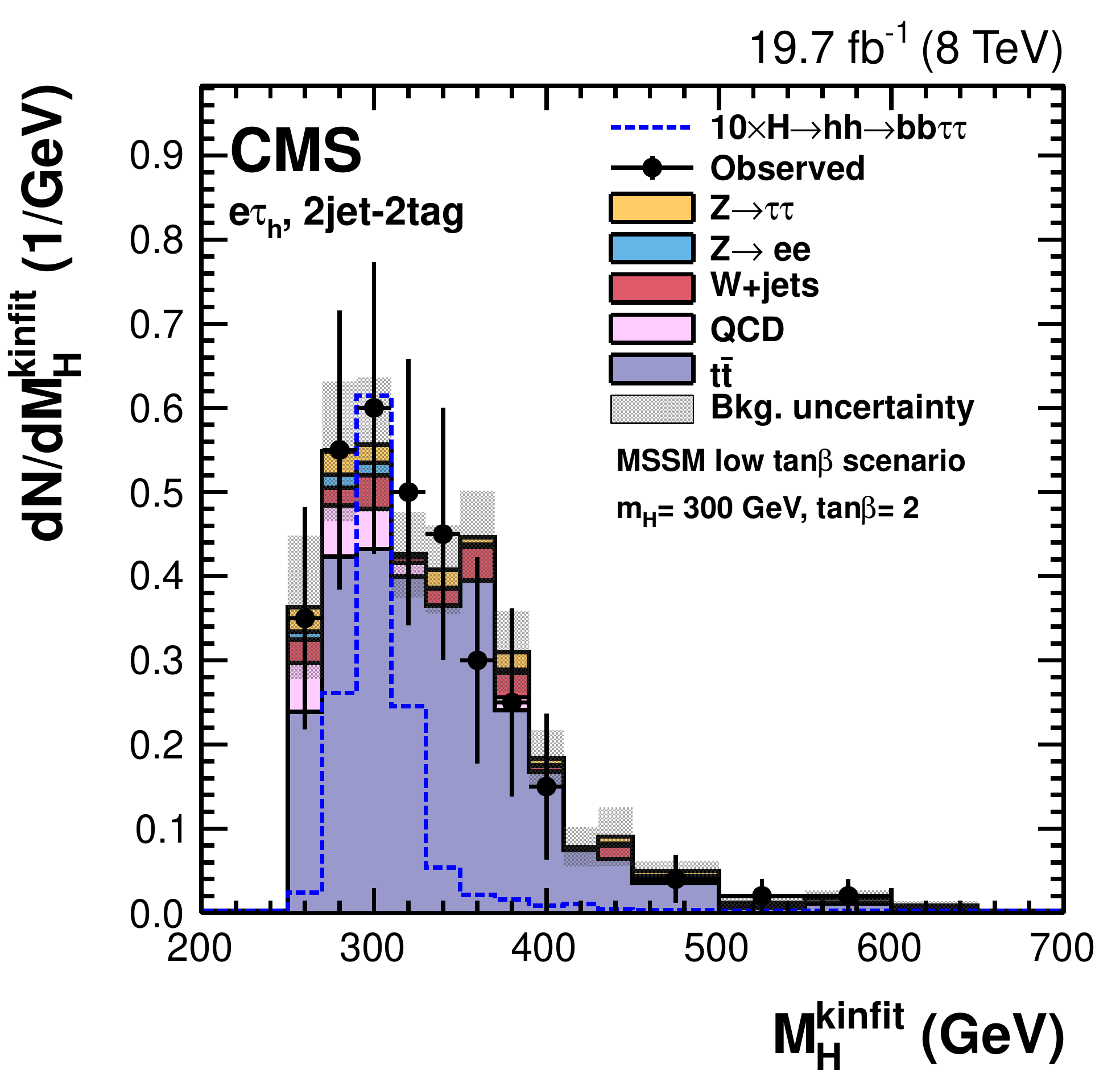}
\caption{
  Distributions of the reconstructed four-body mass with the
  kinematic fit after applying mass selections on $m_{\Pgt\Pgt}$ and $m_{\cPqb\cPqb}$
  in the $\Pe\tauh$
  channel. The plots are shown for events in the 2jet--0tag (\cmsTopLeft), 2jet--1tag
  (\cmsTopRight), and 2jet--2tag (bottom) categories. The expected signal scaled by a
factor 10 is shown superimposed as an open dashed histogram for $\tan\beta = 2$
and $m_{\PH}= 300$\GeV in the low $\tan\beta$ scenario of the MSSM.
  Expected background contributions are shown for the values of nuisance parameters (systematic uncertainties)
  obtained after fitting the signal plus background hypothesis to the data.
}
\label{fig:resultsKinFitMassCutsETau}
\end{figure}

\begin{figure}[htbp]
\centering
\includegraphics[width=\cmsFigWidthThree]{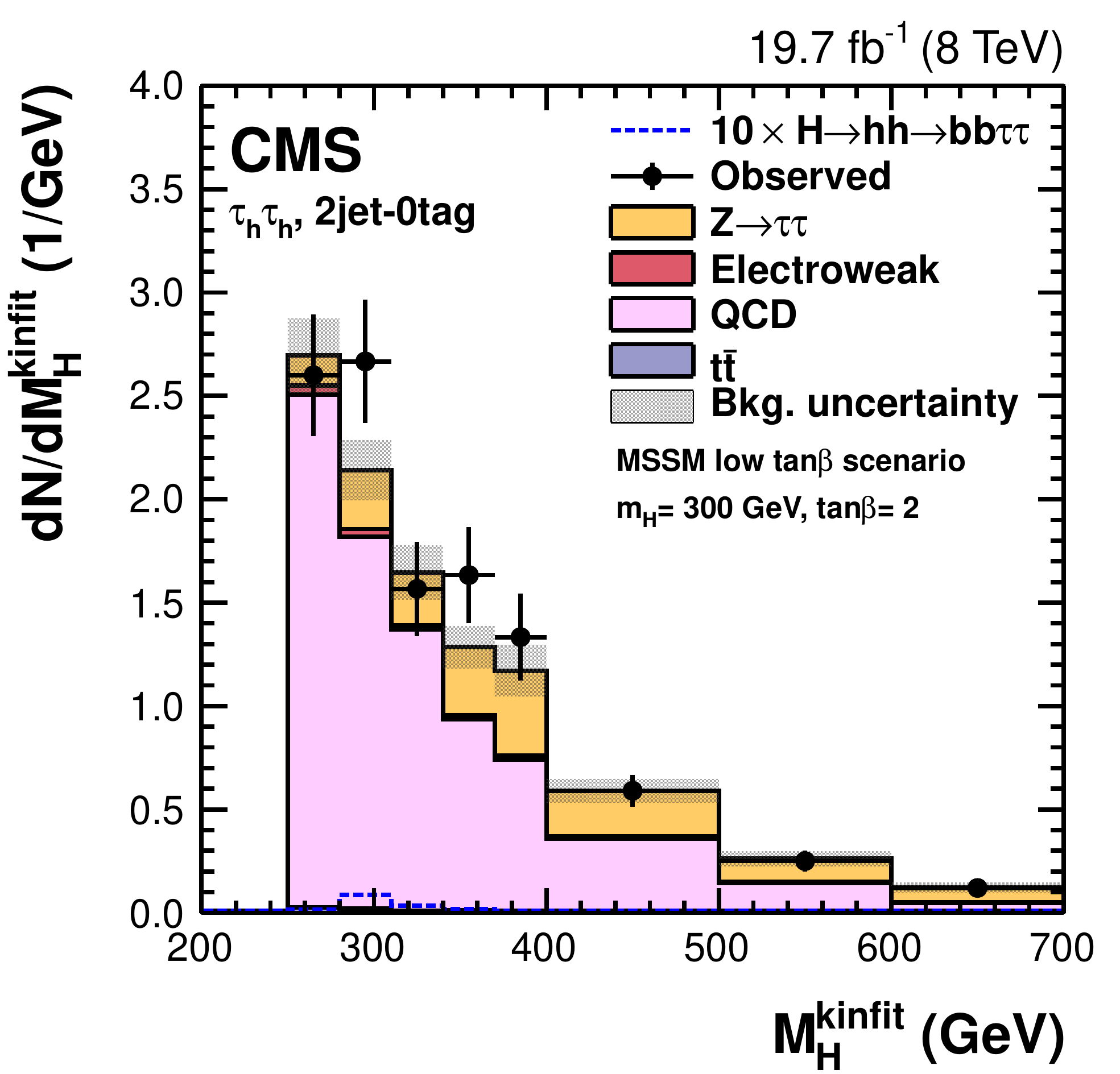}
\includegraphics[width=\cmsFigWidthThree]{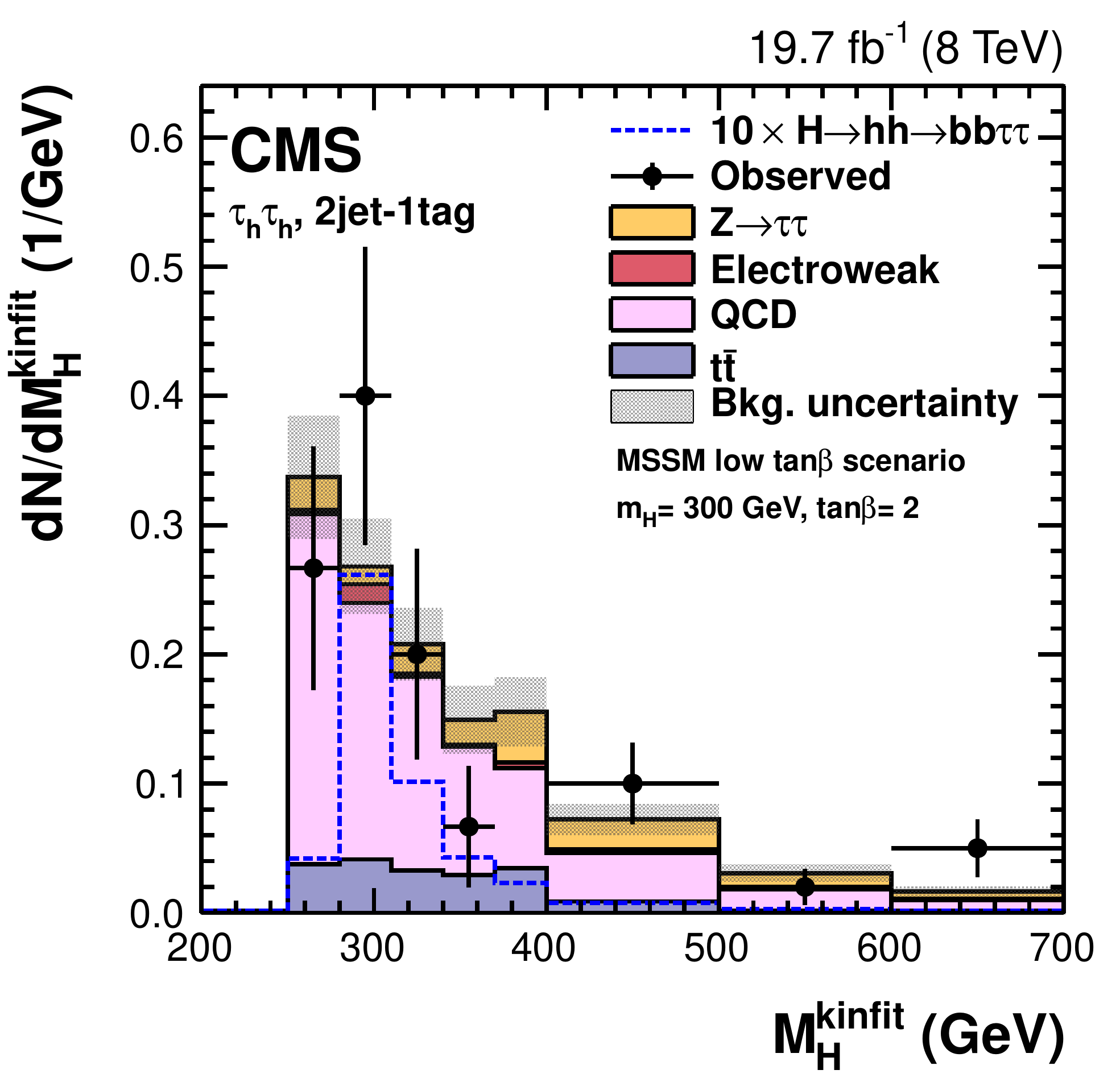}
\includegraphics[width=\cmsFigWidthThree]{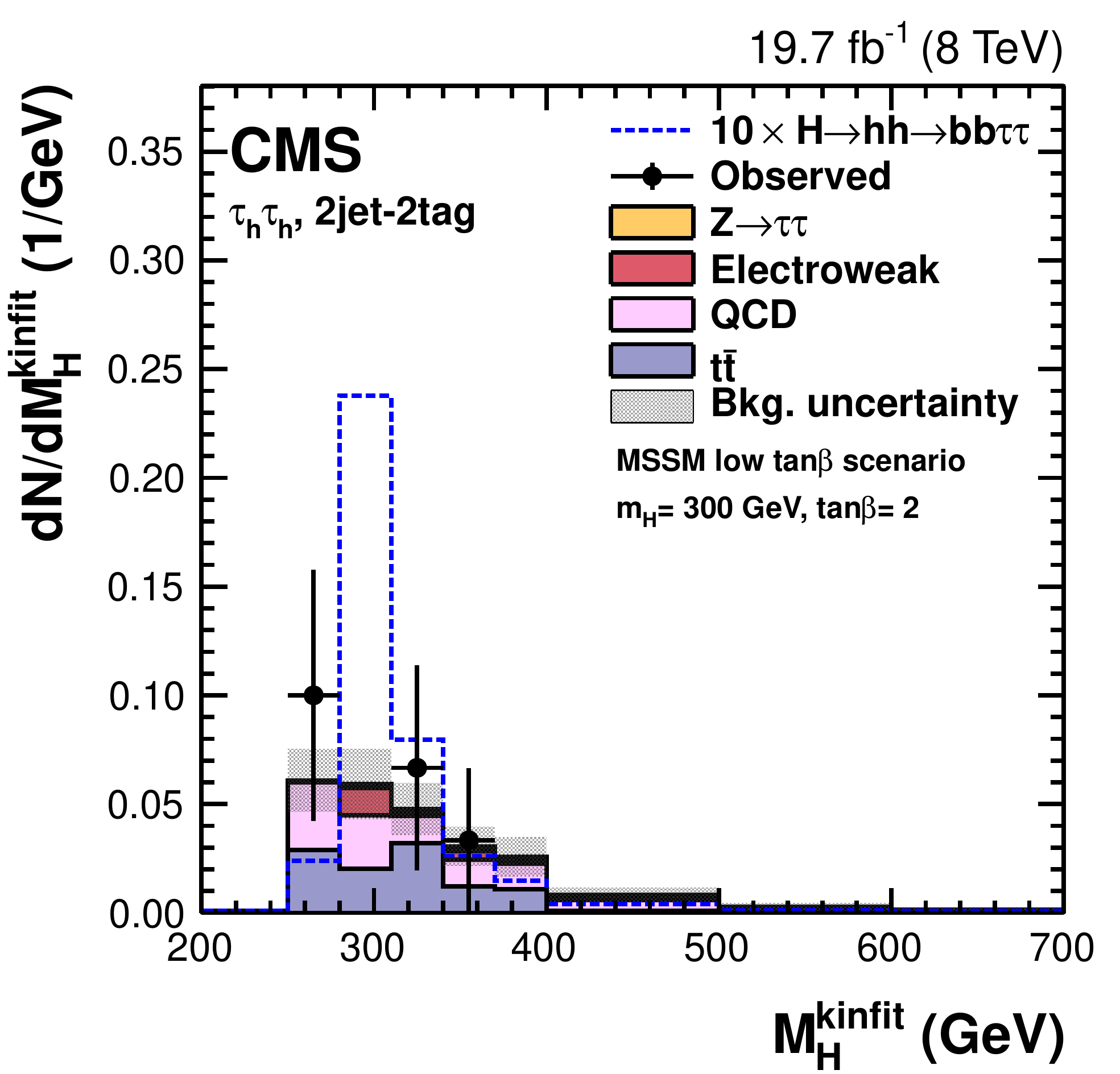}
\caption{
  Distributions of the reconstructed four-body mass with the
  kinematic fit after applying mass selections on $m_{\Pgt\Pgt}$ and $m_{\cPqb\cPqb}$
  in the $\tauh\tauh$
  channel. The plots are shown for events in the 2jet--0tag (\cmsTopLeft), 2jet--1tag
  (\cmsTopRight), and 2jet--2tag (bottom) categories. The expected signal scaled by a
factor 10 is shown superimposed as an open dashed histogram for $\tan\beta = 2$
and $m_{\PH}= 300$\GeV in the low $\tan\beta$ scenario of the MSSM.
  Expected background contributions are shown for the values of nuisance parameters (systematic uncertainties)
  obtained after fitting the signal plus background hypothesis to the data.
}
\label{fig:resultsKinFitMassCutsTauTau}
\end{figure}

The signal-to-background ratio is greatly
improved by selecting events that are consistent with a mass of 125 $\GeV$ for both
the dijet ($m_{\cPqb\cPqb}$) mass and the ditau mass ($m_{\Pgt\Pgt}$) reconstructed with \textsc{SVFit}. The mass windows of the selections are
optimised to collect as much signal as possible while rejecting a large
part of the background. They correspond to $70 < m_{\cPqb\cPqb} < 150 \GeV$ and $90 <
m_{\tau\tau} < 150 \GeV$. The invariant mass distributions of the $\PH$ boson in different final states are shown in Figs.~\ref{fig:resultsKinFitMassCutsMuTau}, \ref{fig:resultsKinFitMassCutsETau} and~\ref{fig:resultsKinFitMassCutsTauTau}.

\begin{figure*}[htbp]
\centering
\includegraphics[width=0.49\textwidth]{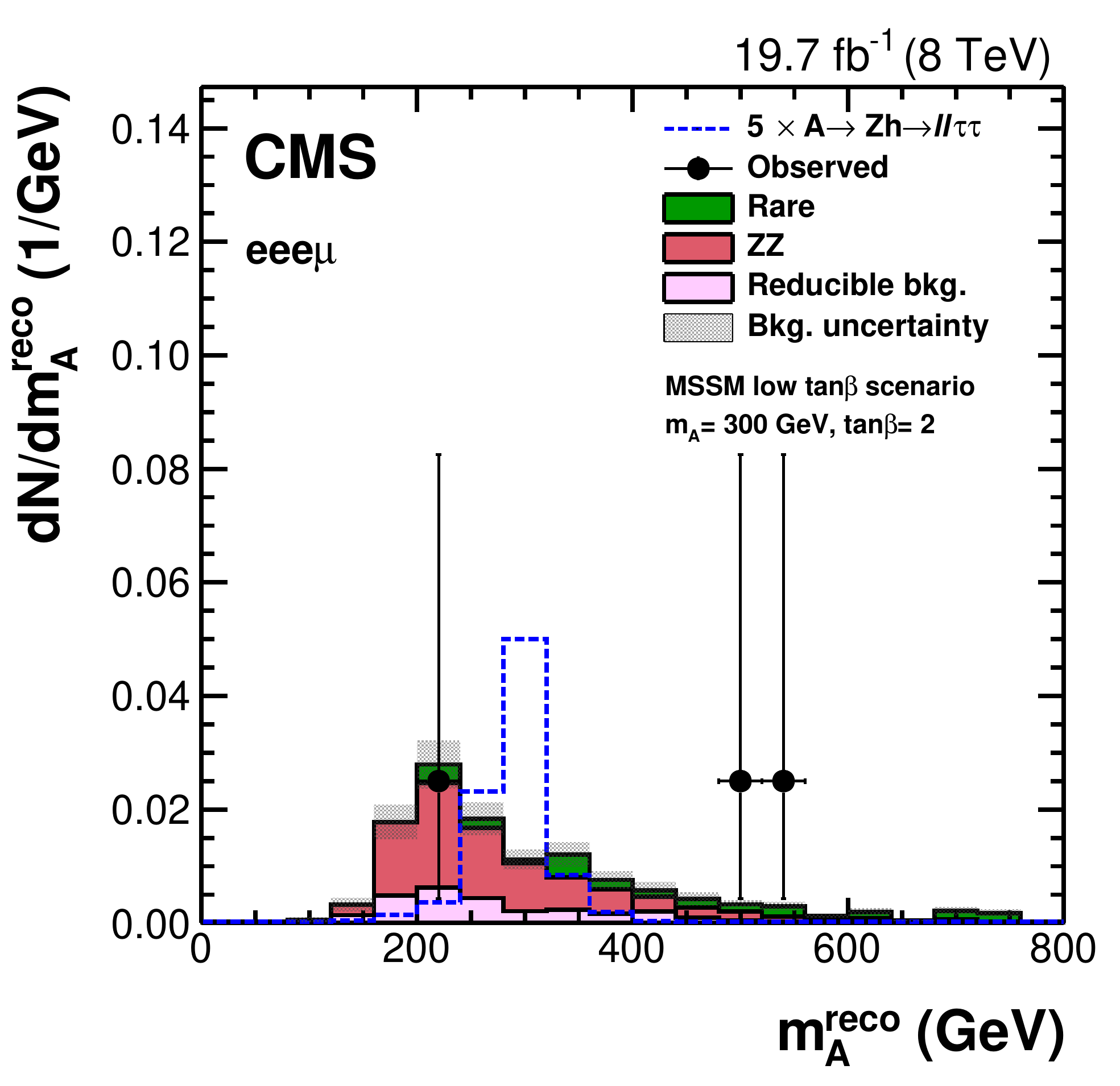}
\includegraphics[width=0.49\textwidth]{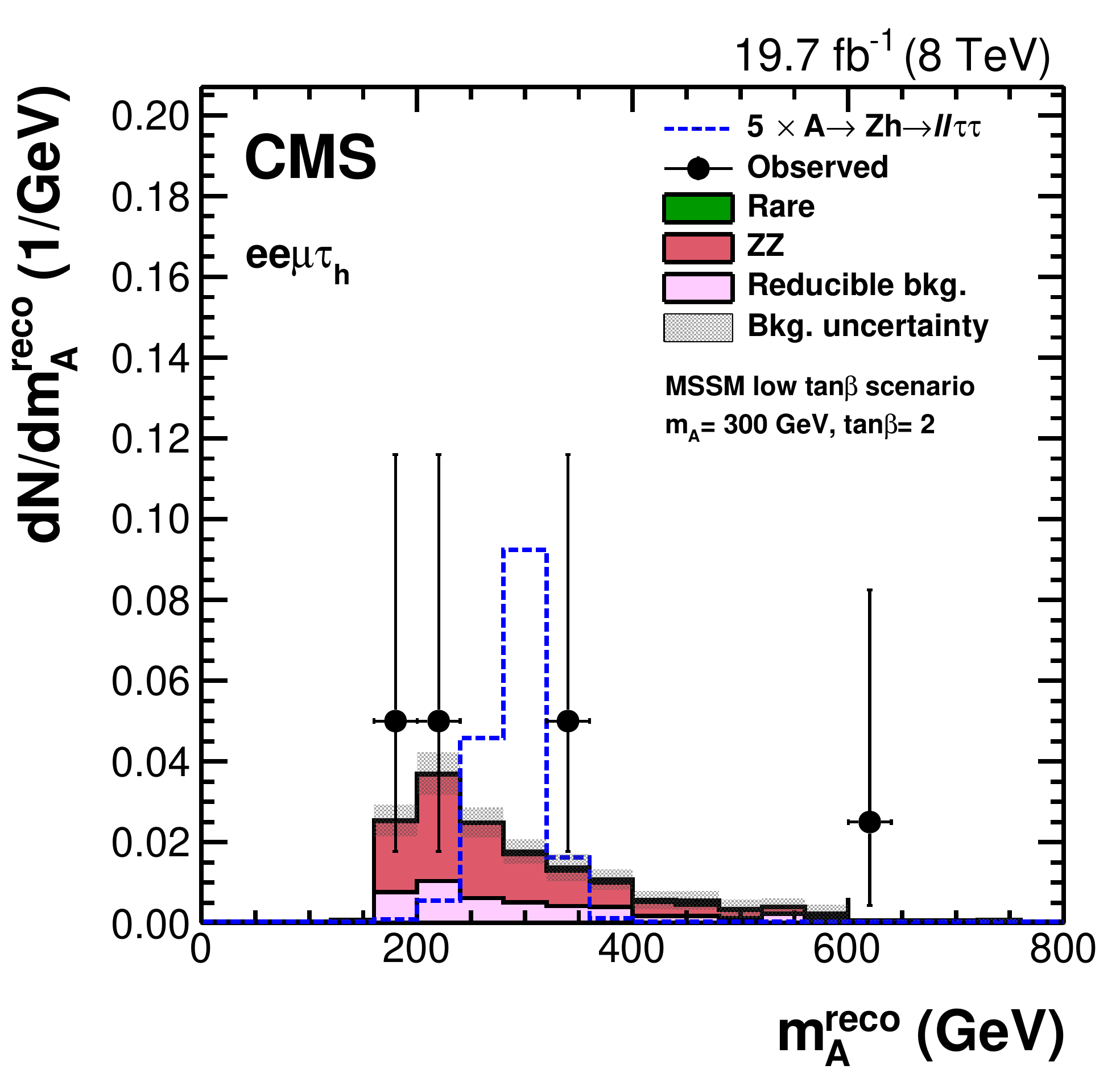}
\includegraphics[width=0.49\textwidth]{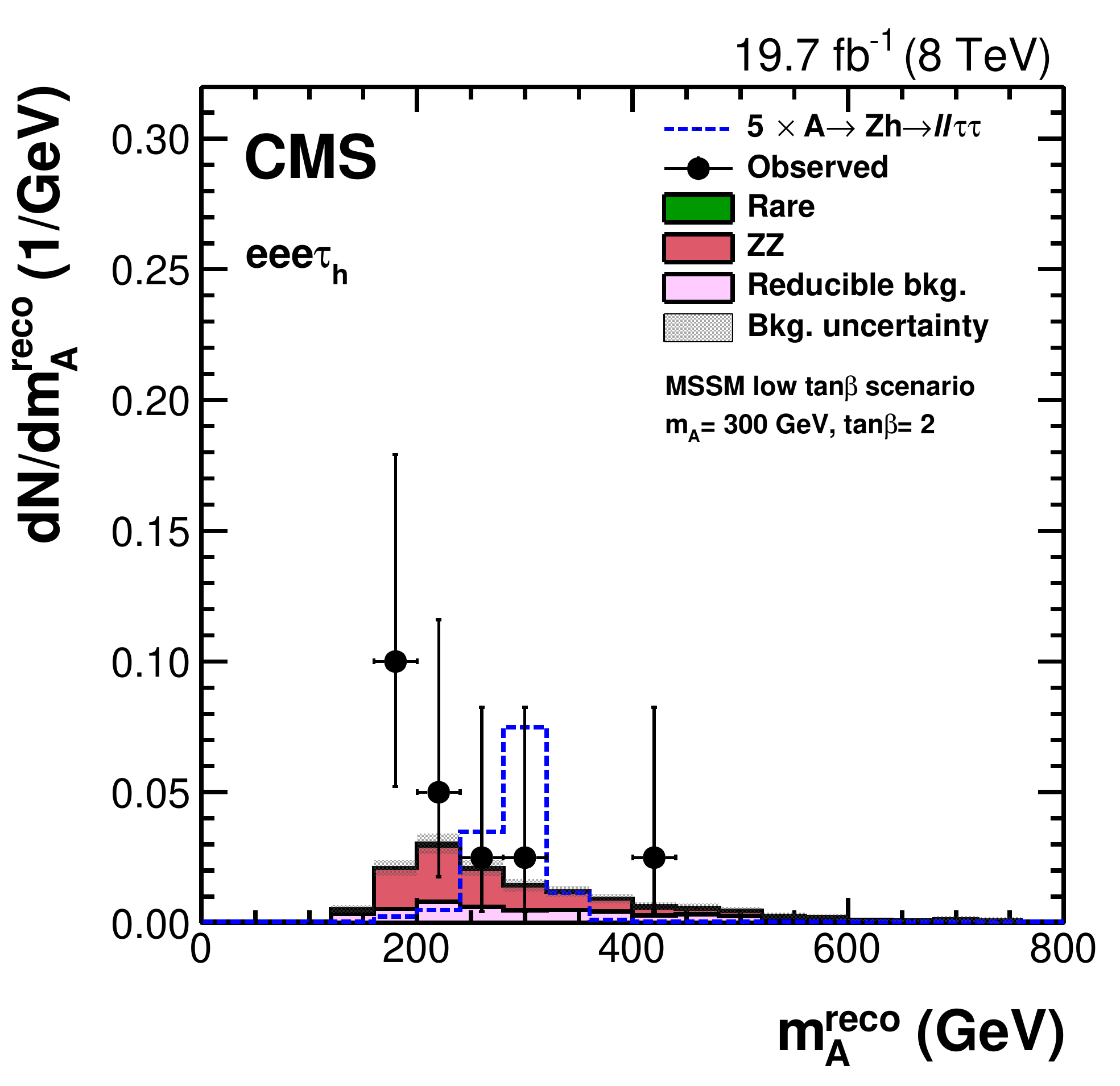}
\includegraphics[width=0.49\textwidth]{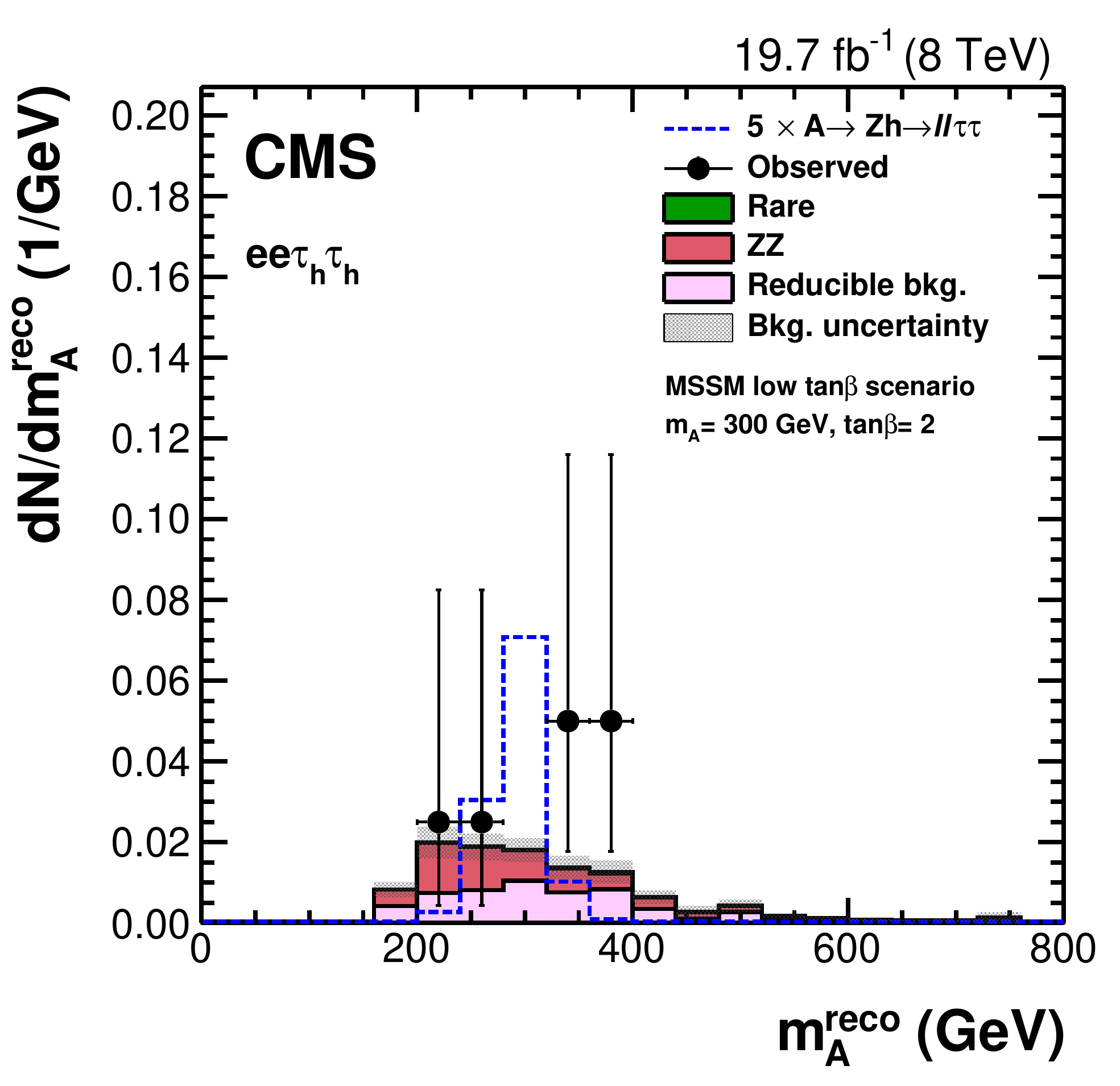}
\caption{Invariant mass distributions for different final states of the
$\PA\to\PZ\Ph$ process where $\PZ$ decays to $\Pe\Pe$. The expected signal scaled by a
factor 5 is shown superimposed as an open dashed histogram for $\tan\beta = 2$
and $m_{\PA}= 300$\GeV in the low $\tan\beta$ scenario of MSSM.
Expected background contributions are shown for the values of nuisance parameters (systematic uncertainties)
obtained after fitting the signal plus background hypothesis to the data.}
\label{fig:massplots_A_Zee}
\end{figure*}

\begin{figure*}[htbp]
\centering
\includegraphics[width=0.49\textwidth]{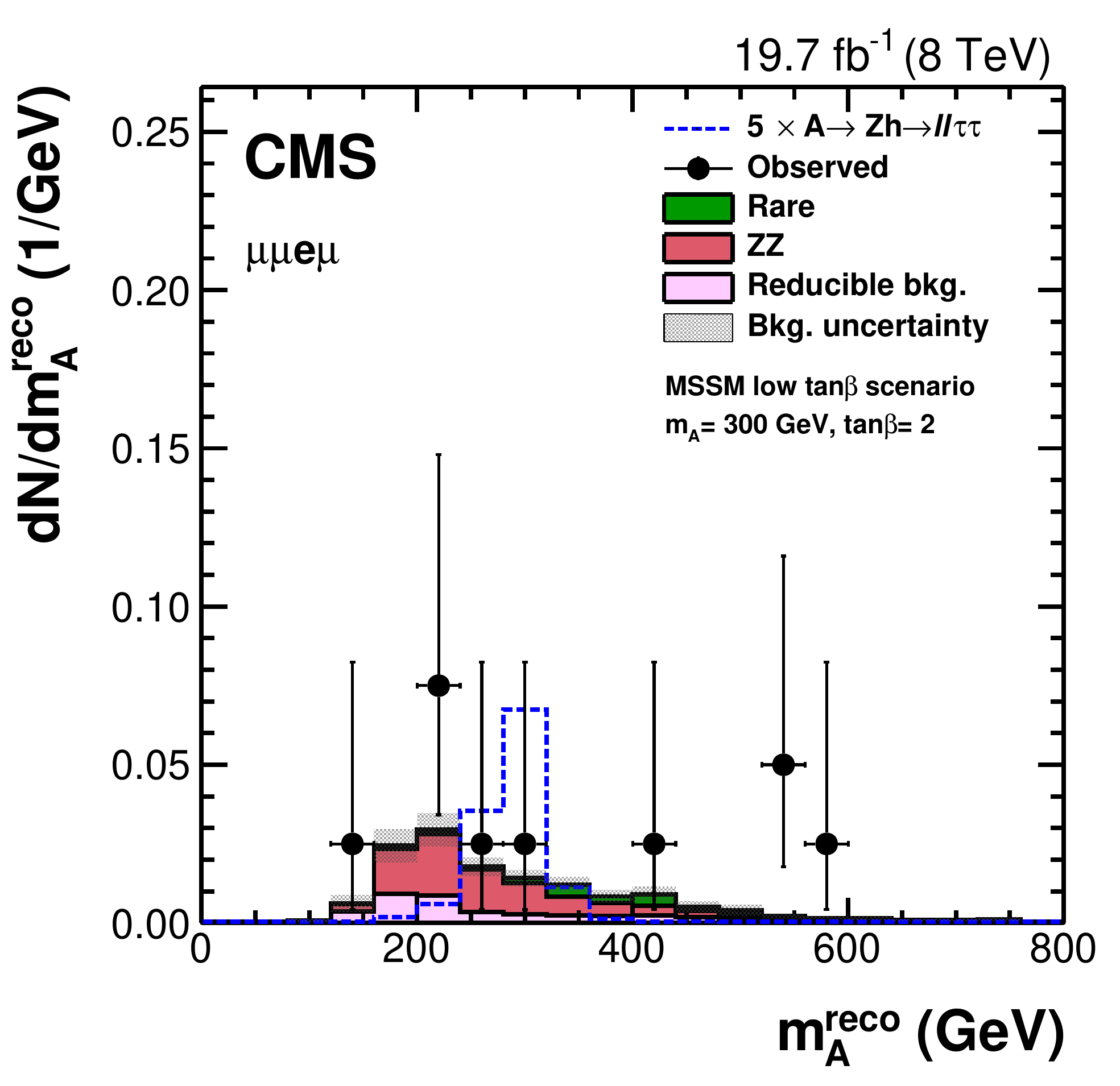}
\includegraphics[width=0.49\textwidth]{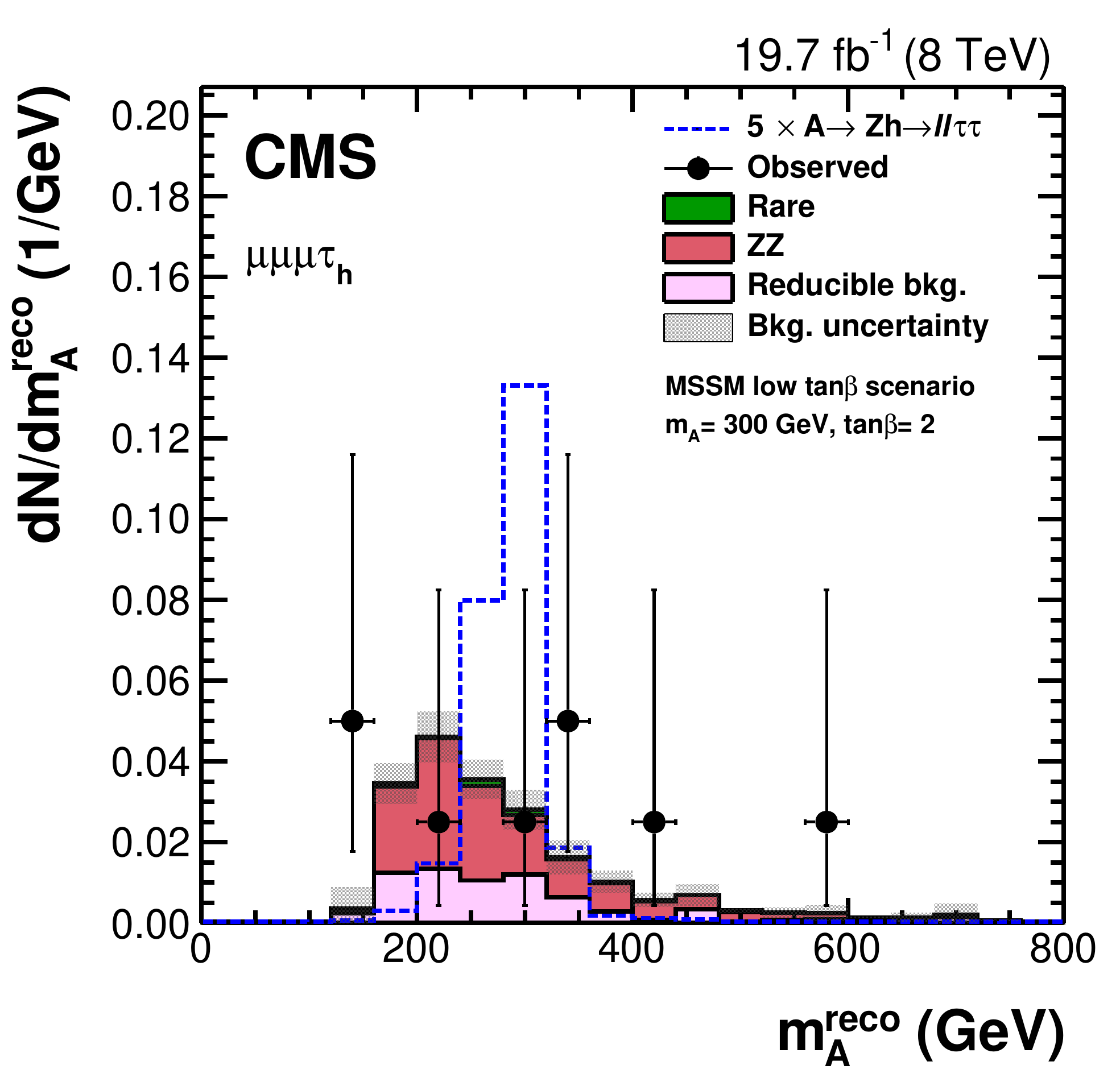}
\includegraphics[width=0.49\textwidth]{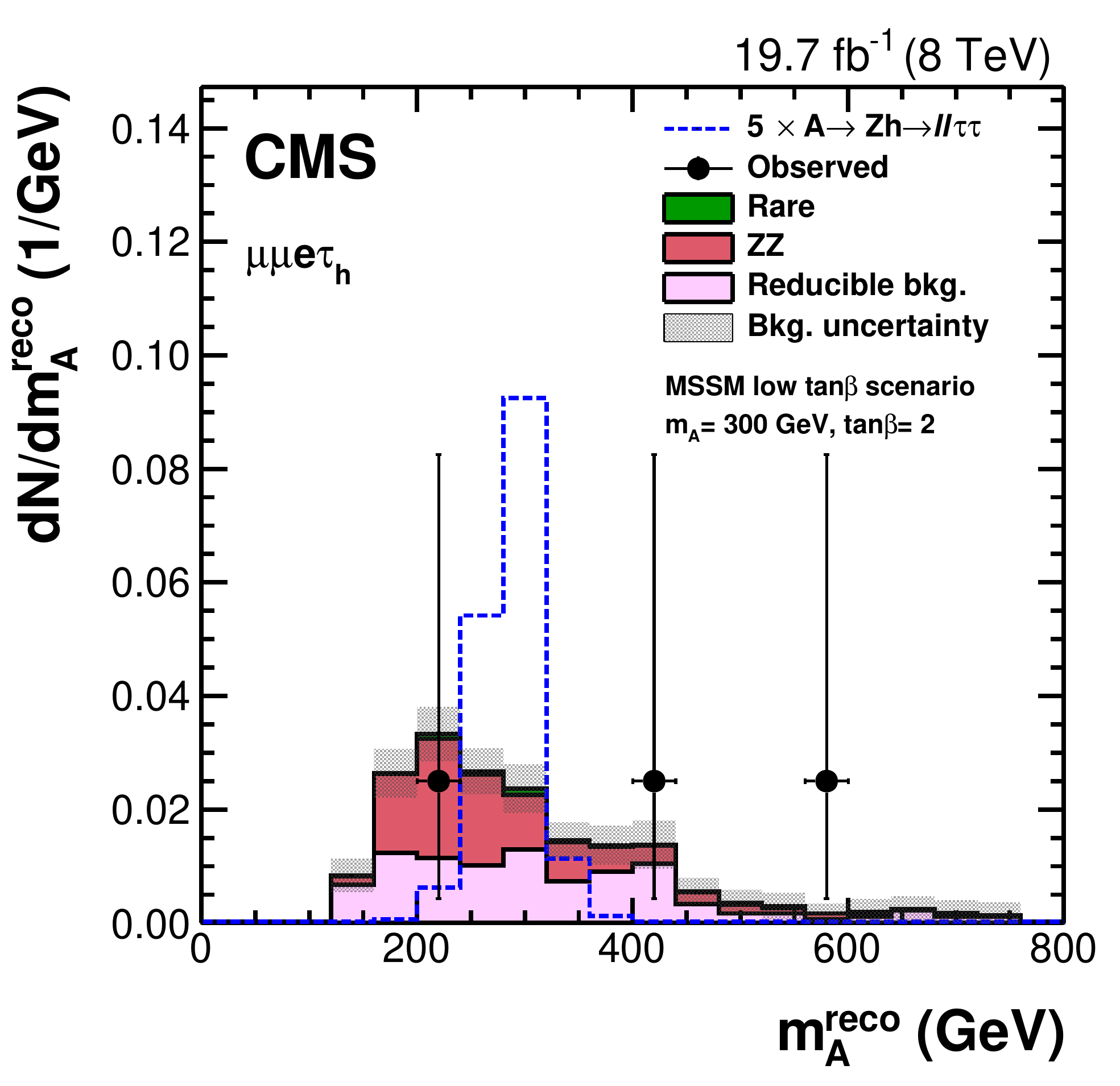}
\includegraphics[width=0.49\textwidth]{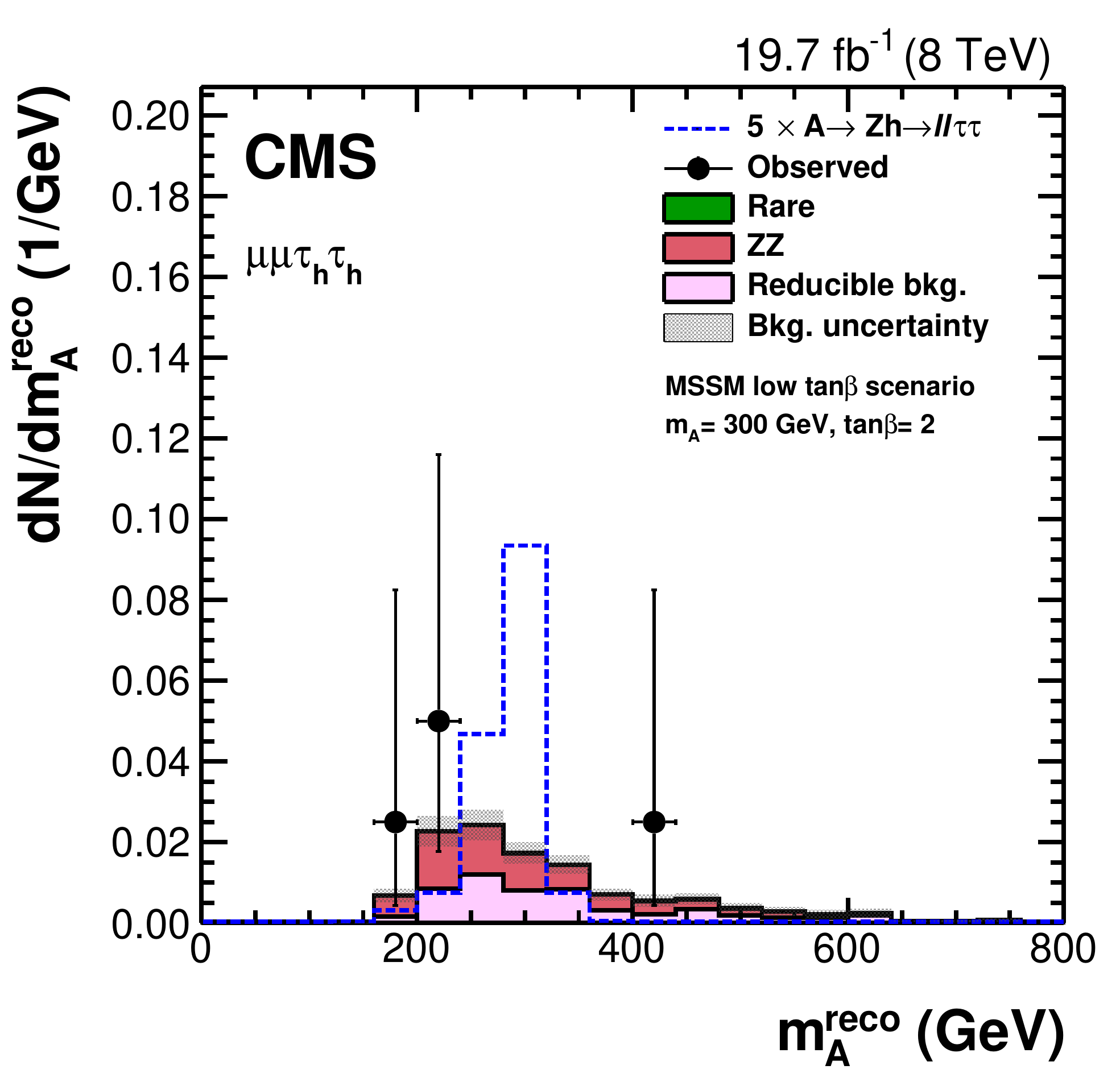}
\caption{Invariant mass distributions for different final states of the
$\PA\to\PZ\Ph$ process where $\PZ$ decays to $\Pgm\Pgm$. The expected signal scaled by a
factor 5 is shown superimposed as an open dashed histogram for $\tan\beta = 2$
and $m_{\PA}= 300$\GeV in the low $\tan\beta$ scenario of MSSM.
Expected background contributions are shown for the values of nuisance parameters (systematic uncertainties)
obtained after fitting the signal plus background hypothesis to the data.}
\label{fig:massplots_A_Zmumu}
\end{figure*}

For the $\PA\to\PZ\Ph\to{\ell\ell}\Pgt\Pgt$ process, the $\PA$ boson mass is
reconstructed from the four-vector information of the $\PZ$ boson candidate and the four-vector information of the $\Ph$ boson candidate as obtained from \textsc{SVFit}.
The invariant mass distributions of the $\PA$ boson in the different final
states are shown in
Figs.~\ref{fig:massplots_A_Zee} and~\ref{fig:massplots_A_Zmumu}.
The $\ell\ell\tauh\tauh$ final states have a comparable contribution from reducible and irreducible backgrounds, while the $\ell\ell\Pe\Pgm$ final states are
dominated by the irreducible $\PZ\PZ$ production. The background in labelled as
``rare'' collects together the smaller contributions from the triboson processes
as discussed in the previous section.

In neither search do the invariant mass spectra show any evidence of a signal.
Model independent upper limits at 95\% confidence level (CL) on the cross section times branching
fraction are set using a binned maximum likelihood fit for the
\emph{signal plus background} and \emph{background--only} hypotheses.
The limits are determined using the \CLs method~\cite{Junk:1999kv,Read:2002hq} and the procedure is described in Ref. ~\cite{LHC-HCG-Report, Cowan:2010js}.

Systematic uncertainties are taken into account as nuisance parameters in the fit procedure: normalisation uncertainties affect the signal and background
yields. Uncertainties on the $\Pgt$ energy scale and jet energy scale are
propagated as shape uncertainties.

\begin{figure*}[htbp]
\centering
\includegraphics[width=0.49\textwidth]{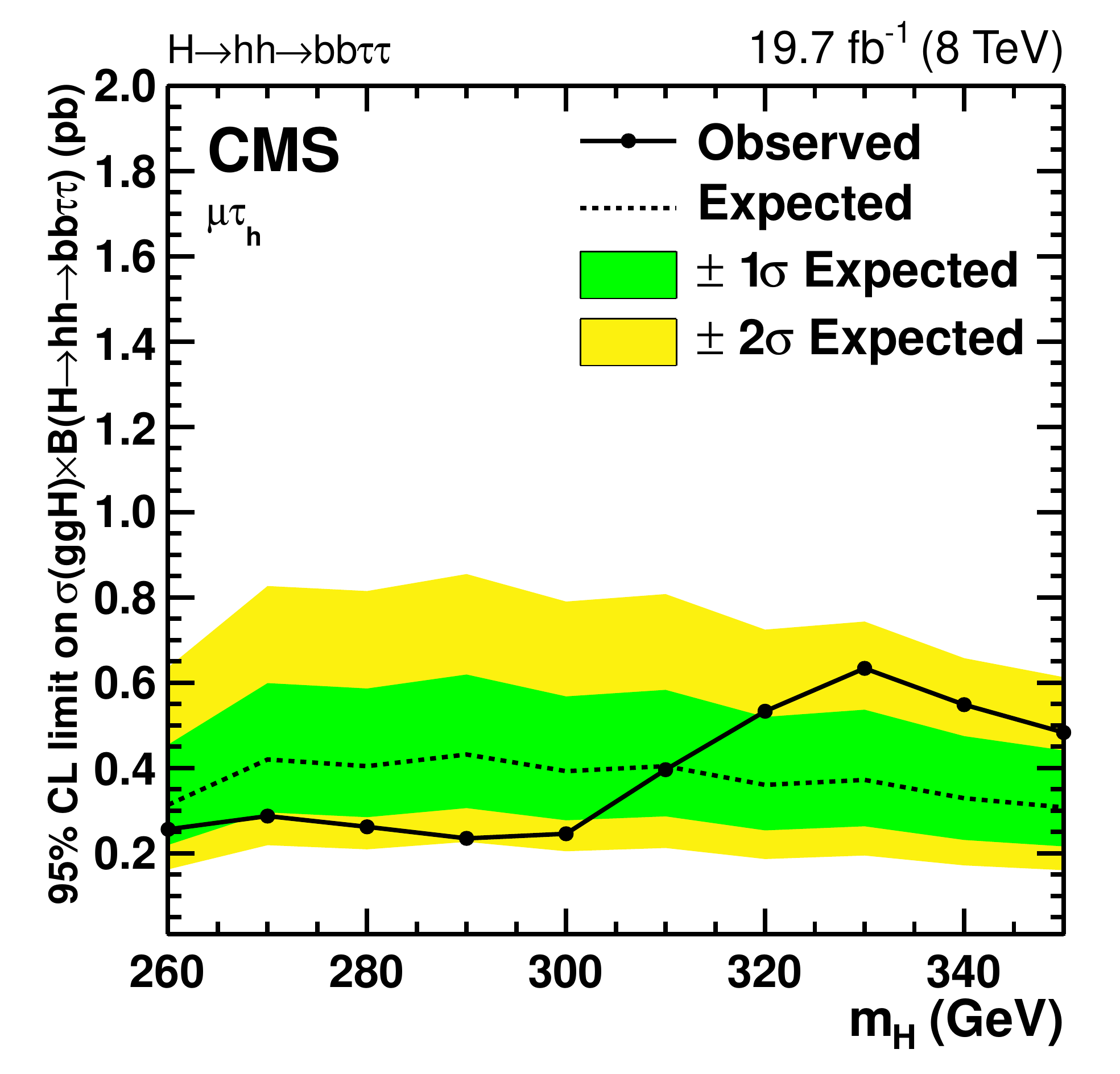}
\includegraphics[width=0.49\textwidth]{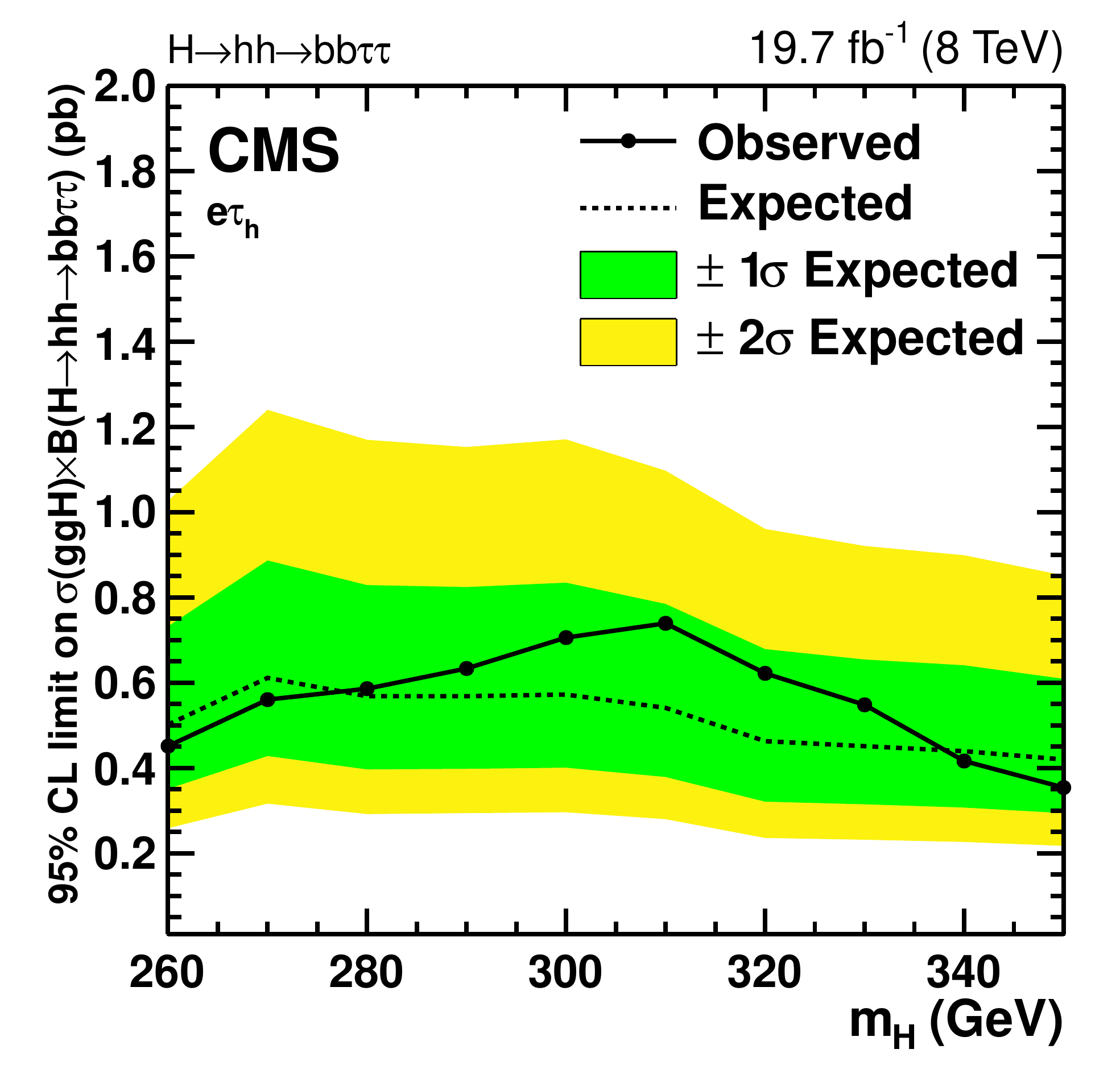}
\includegraphics[width=0.49\textwidth]{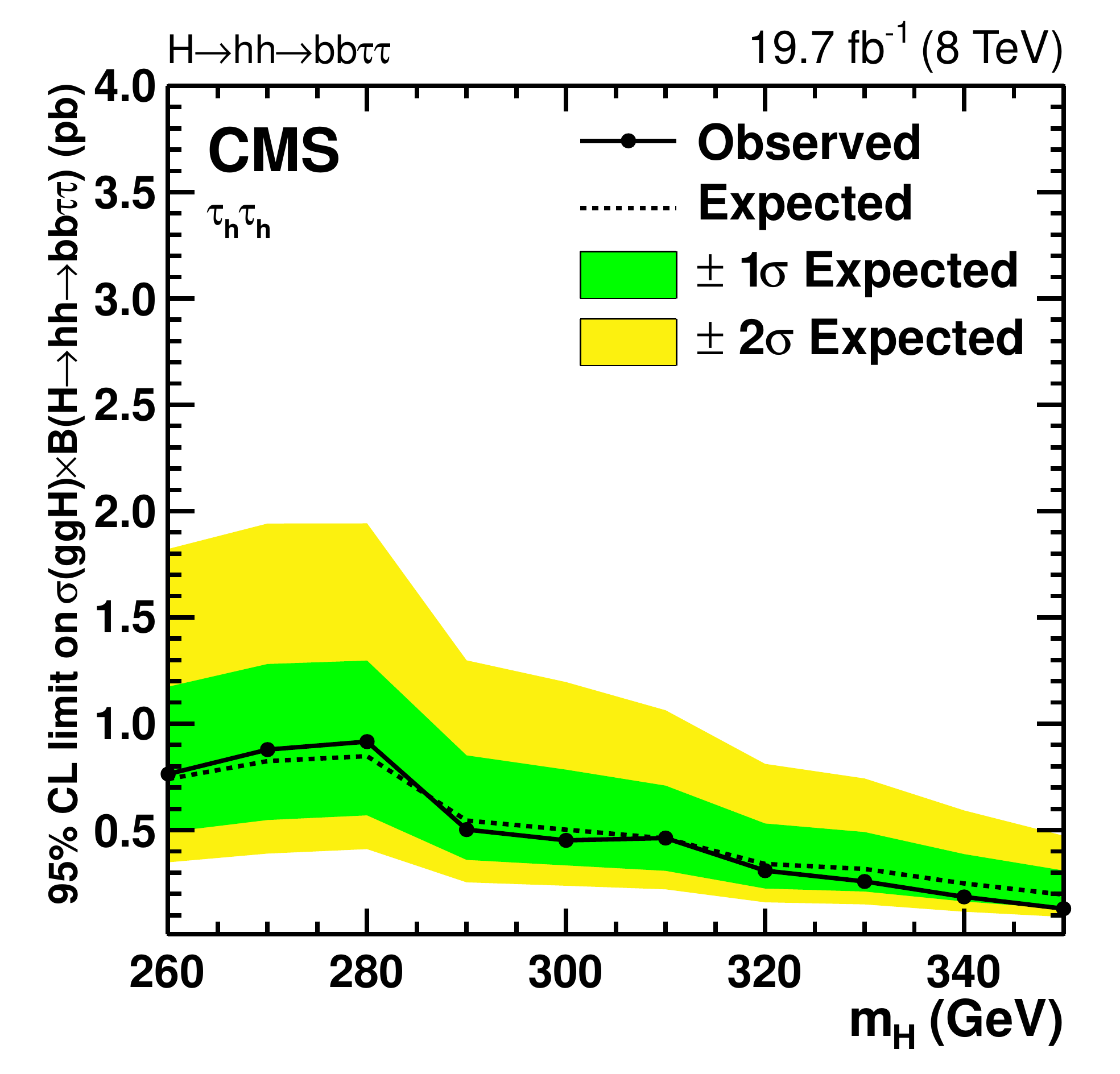}
\includegraphics[width=0.49\textwidth]{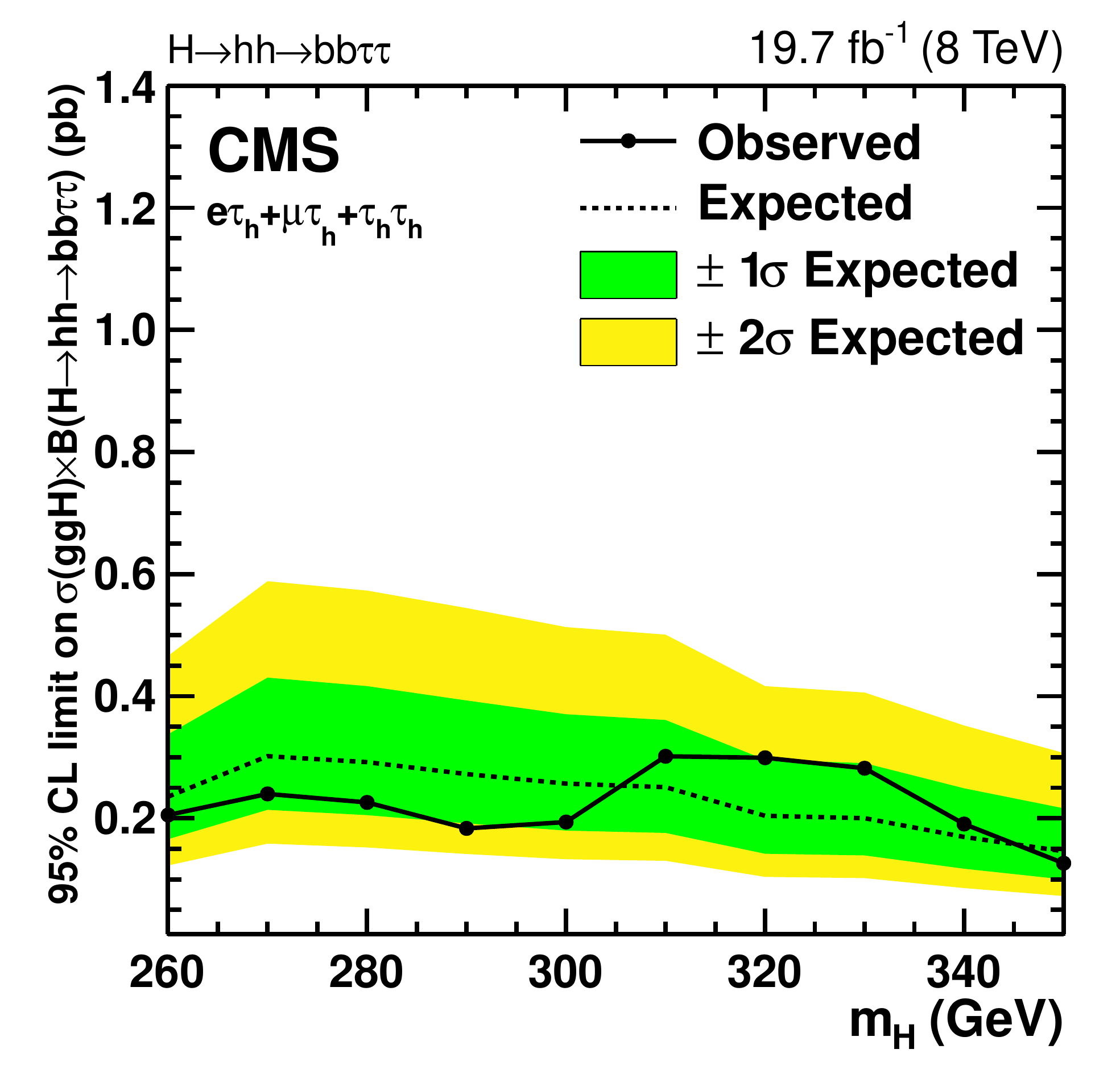}
\caption{
   Upper limits  at 95\% CL on the $\PH\to\Ph\Ph\to\cPqb\cPqb\Pgt\Pgt$ cross section times branching fraction for the $\Pgm\tauh$ (top left), $\Pe\tauh$ (top right), $\tauh\tauh$ (bottom left), and for final states combined (bottom right)
}
\label{fig:results_ggHTohh_indivChannels}
\end{figure*}

\begin{figure*}[htbp]
\centering
\includegraphics[width=0.49\textwidth]{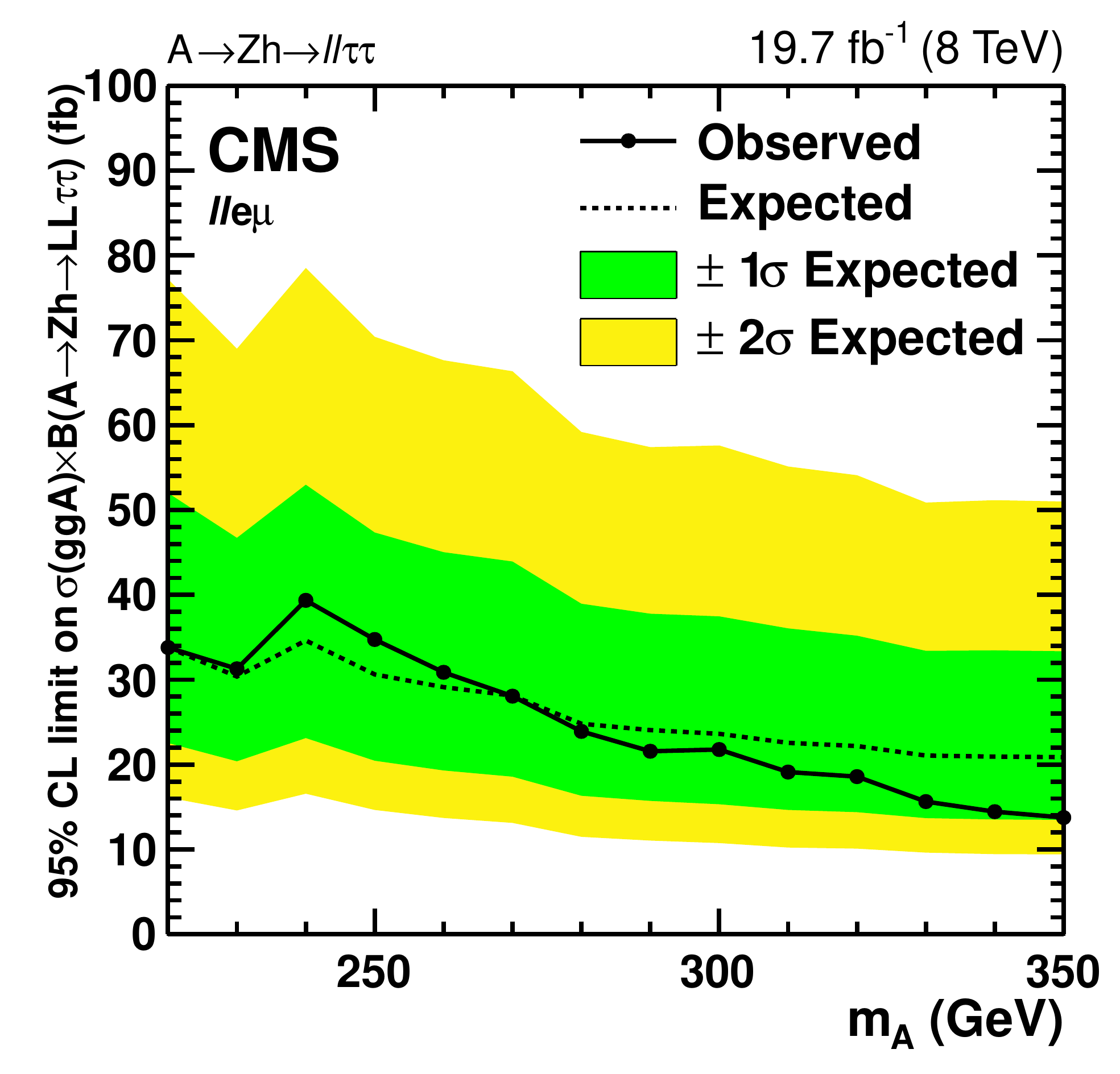}
\includegraphics[width=0.49\textwidth]{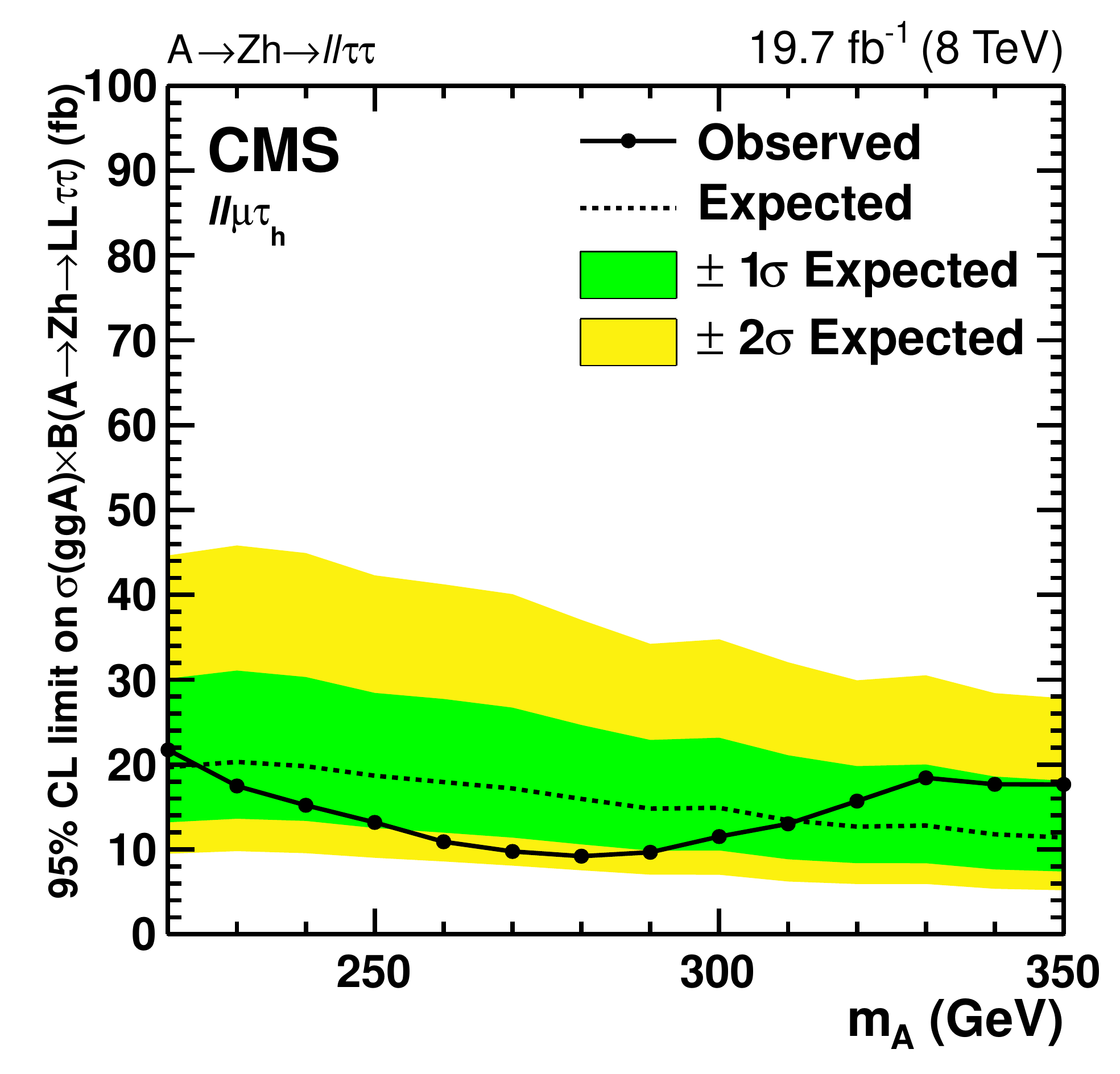}
\includegraphics[width=0.49\textwidth]{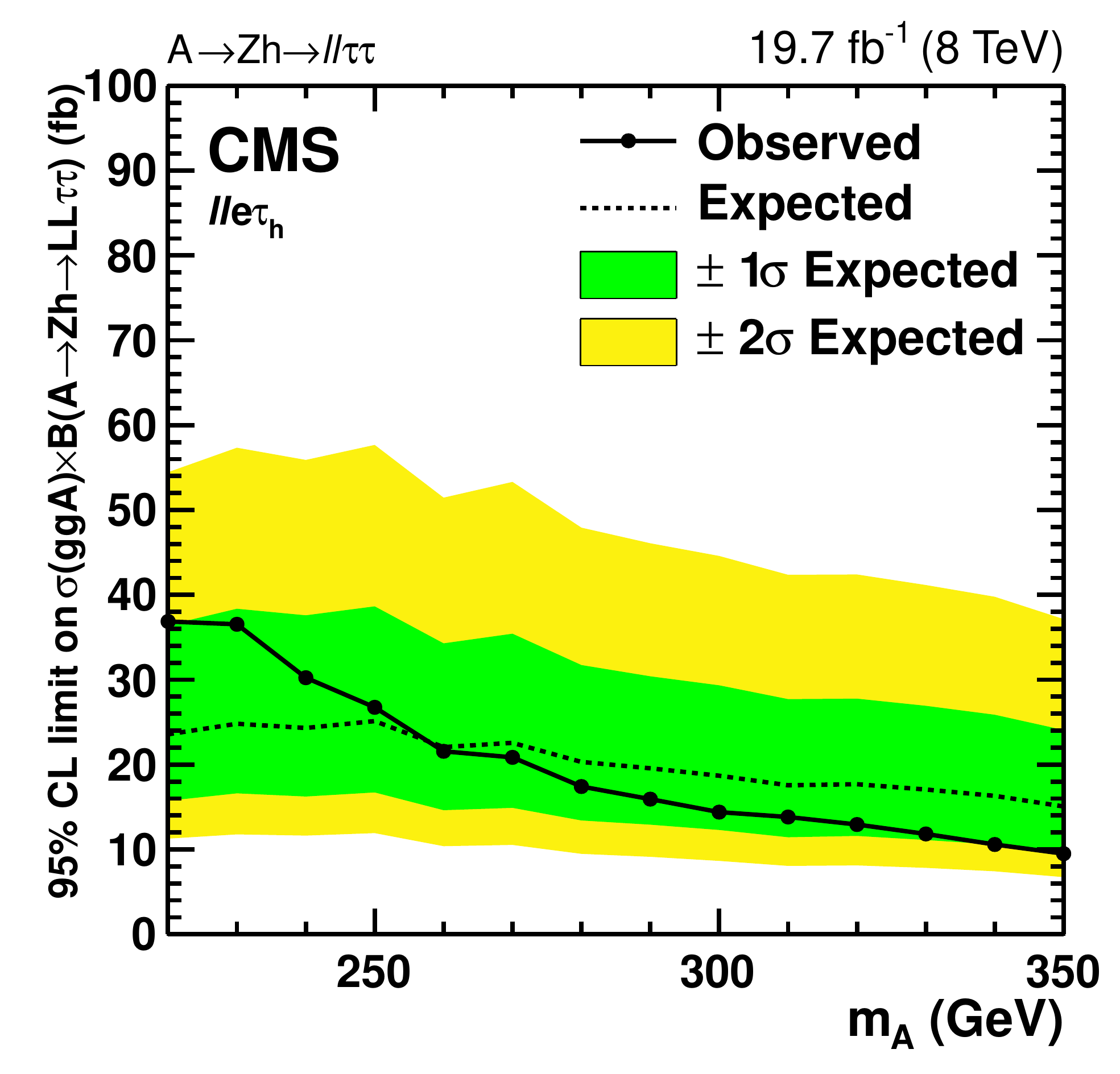}
\includegraphics[width=0.49\textwidth]{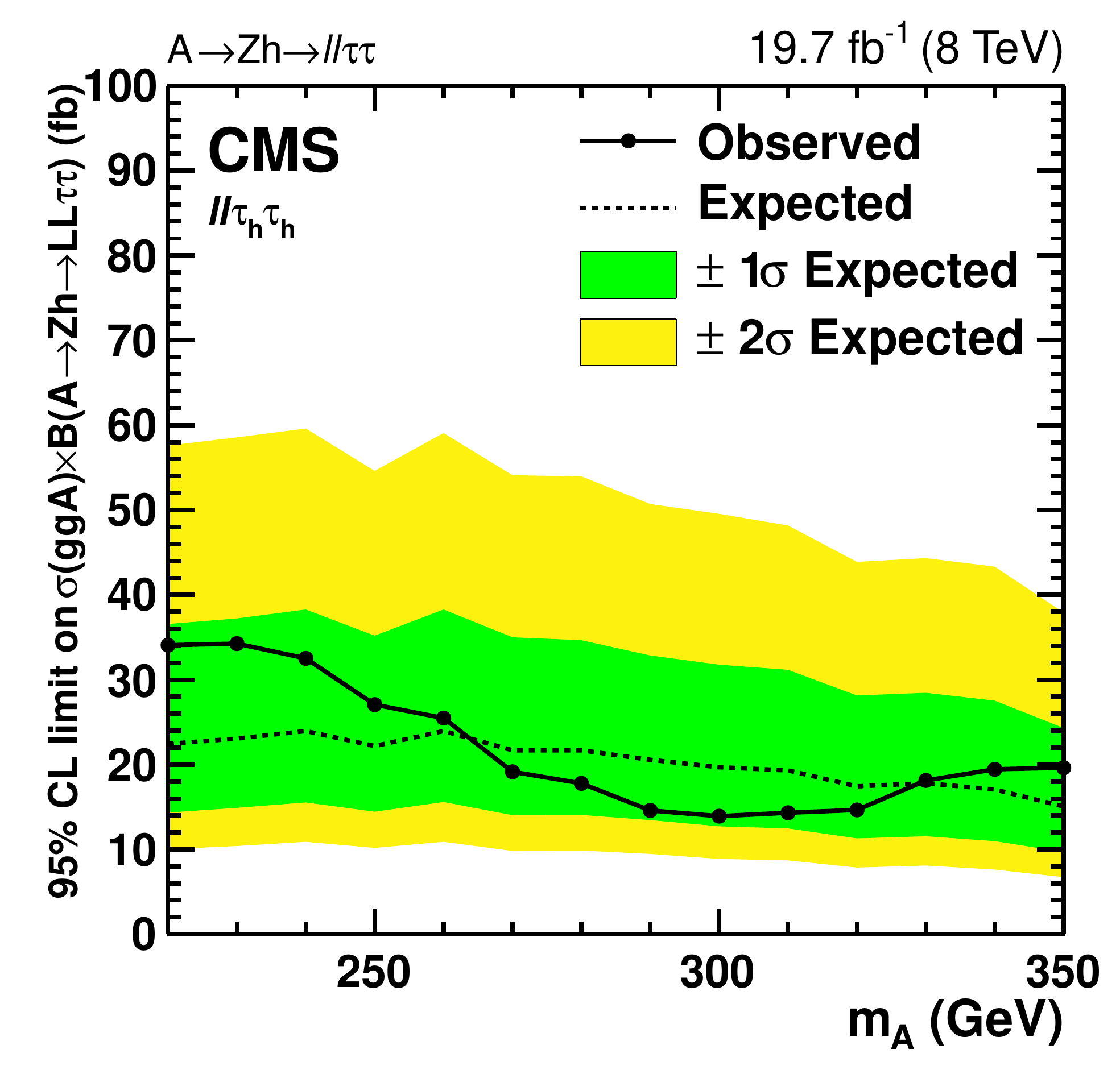}
\caption{Upper limits at 95\% CL on cross section times branching fraction on
$\PA\to\PZ\Ph\to LL\Pgt\Pgt$ for $\ell\ell\Pe\Pgm$ (top left), $\ell\ell\Pgm\tauh$ (top right), $\ell\ell\Pe\tauh$ (bottom left), and $\ell\ell\tauh\tauh$ (bottom right) final states.
}
\label{fig:limits_chl_A}
\end{figure*}

\begin{figure}[htb]
\centering
\includegraphics[width=0.49\textwidth]{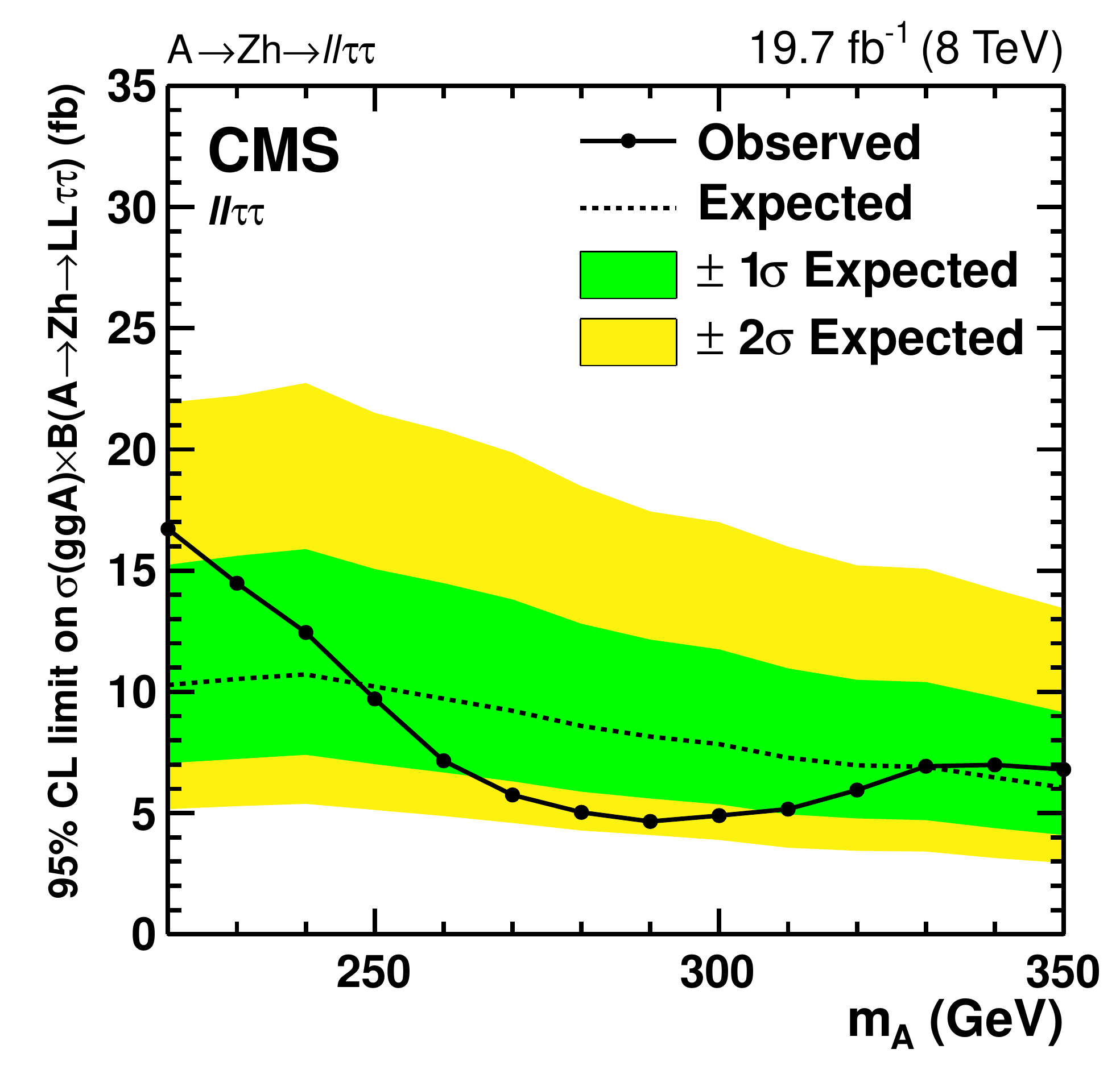}
\includegraphics[width=0.49\textwidth]{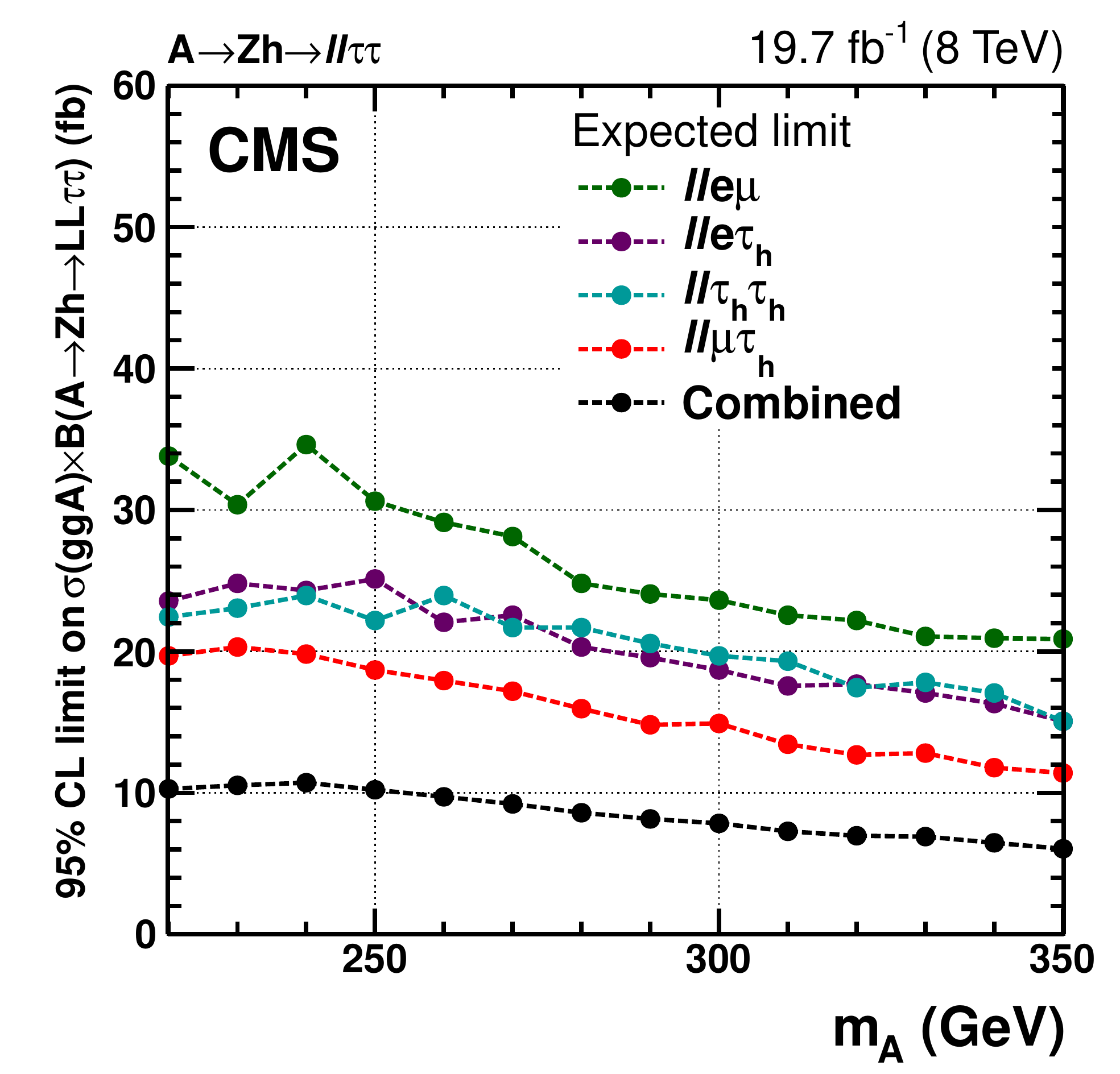}
\caption{Upper limits at 95\% CL on cross section times branching fraction on
$\PA\to\PZ\Ph\to LL \Pgt\Pgt$ for all ${\ell\ell}\Pgt\Pgt$ final states combined (\cmsLeft) and comparison of the different final states (\cmsRight).
}
\label{fig:limits_all_A}
\end{figure}

The model independent expected and observed cross section times branching fraction limits for the $\PH\to\Ph\Ph\to\cPqb\cPqb\Pgt\Pgt$ process are shown
in Fig.~\ref{fig:results_ggHTohh_indivChannels} and for the $\PA\to\PZ\Ph\to LL\Pgt\Pgt$ process in Figs.~\ref{fig:limits_chl_A} and \ref{fig:limits_all_A} where
$L=\Pe,\Pgm$ or $\Pgt$ in order to reflect the small $\PZ\to\Pgt\Pgt$  contribution to the signal acceptance.

We interpret the observed limits on the cross section times branching fraction in the MSSM and 2HDM frameworks, discussed in Section~\ref{sec:introduction}.

\begin{figure}[htbp]
\centering
\includegraphics[width=\cmsFigWidth]{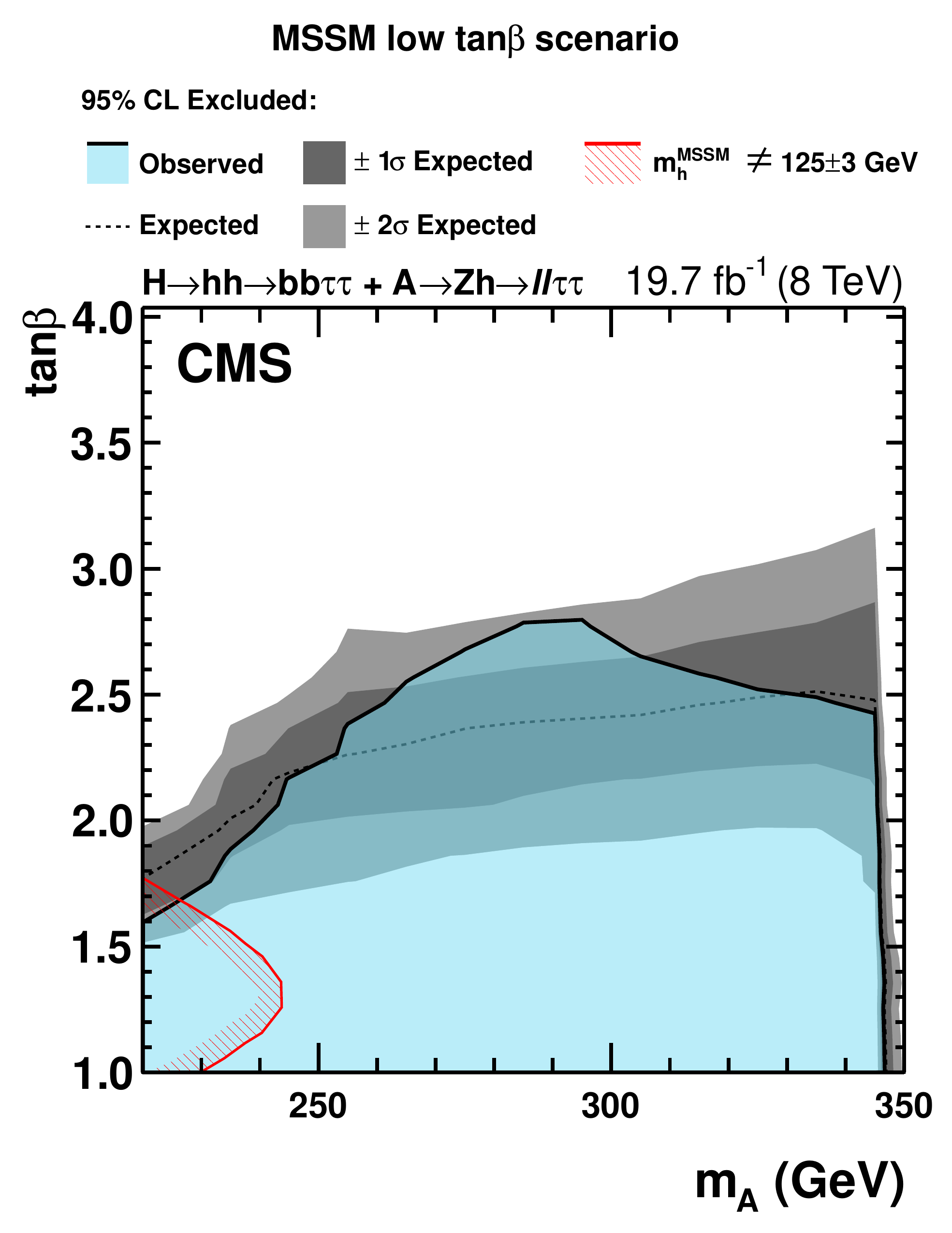}
\caption{The 95\% CL exclusion region in the $m_\PA$-$\tan\beta$ plane for the
  low-$\tan\beta$ scenario as discussed in the introduction, combining the results of
  the $\PH\to\Ph\Ph\to\cPqb\cPqb\Pgt\Pgt$ and the
  $\PA\to\PZ\Ph\to{\ell\ell}\Pgt\Pgt$ analysis. The area highlighted in blue below the black curve
  marks the observed exclusion. The dashed curve and the grey bands show the expected exclusion limit with the relative uncertainty.  The red area with the back-slash
  lines at the lower-left corner of the plot indicates the region excluded by the mass of the SM-like scalar boson being 125\GeV. The limit falls off rapidly as $m_\PA$ approaches
  350\GeV because
decays of the $\PA$ to two top quarks are becoming kinematically allowed.
}
\label{fig:results_lowtanbcmb}
\end{figure}

In the MSSM we interpret them in the ``low tan$\beta$'' scenario~\cite{LHCHXSWGnote,Harlander:2012pb}
in which the value of $M_{\mathrm{SUSY}}$ is increased until the mass of the lightest Higgs boson is
consistent with 125\GeV over a range of low $\tan\beta$ and $m_\PA$
values. The exclusion region in the $m_\PA$-$\tan\beta$ plane  for the combination of the $\PH\to\Ph\Ph\to\cPqb\cPqb\Pgt\Pgt$
and $\PA\to\PZ\Ph\to{\ell\ell}\Pgt\Pgt$ analyses, in such a scenario,
is shown in Fig.~\ref{fig:results_lowtanbcmb}. The limit falls off rapidly as $m_\PA$ approaches 350\GeV because
decays of the $\PA$ to two top quarks are becoming kinematically allowed.

The interpretation of the observed limits in a Type II 2HDM is performed in the ``physics basis''.  The inputs to this interpretation are the
physical Higgs boson masses ($m_{\Ph}$, $m_{\PH}$, $m_{\PA}$, $m_{\PH^{\pm}}$), the ratio of the vacuum
expectation energies ($\tan\beta$), the CP-even Higgs mixing angle ($\alpha$) and $m_{12}^{2}=m_{\PA}^{2}[{\tan\beta}/{(1+\tan\beta^{2})}]$.
For simplicity we assume that $m_{\PH}=m_{\PA}=m_{\PH^{\pm}}$.

\begin{figure}[htbp]
\centering
\includegraphics[width=\cmsFigWidth]{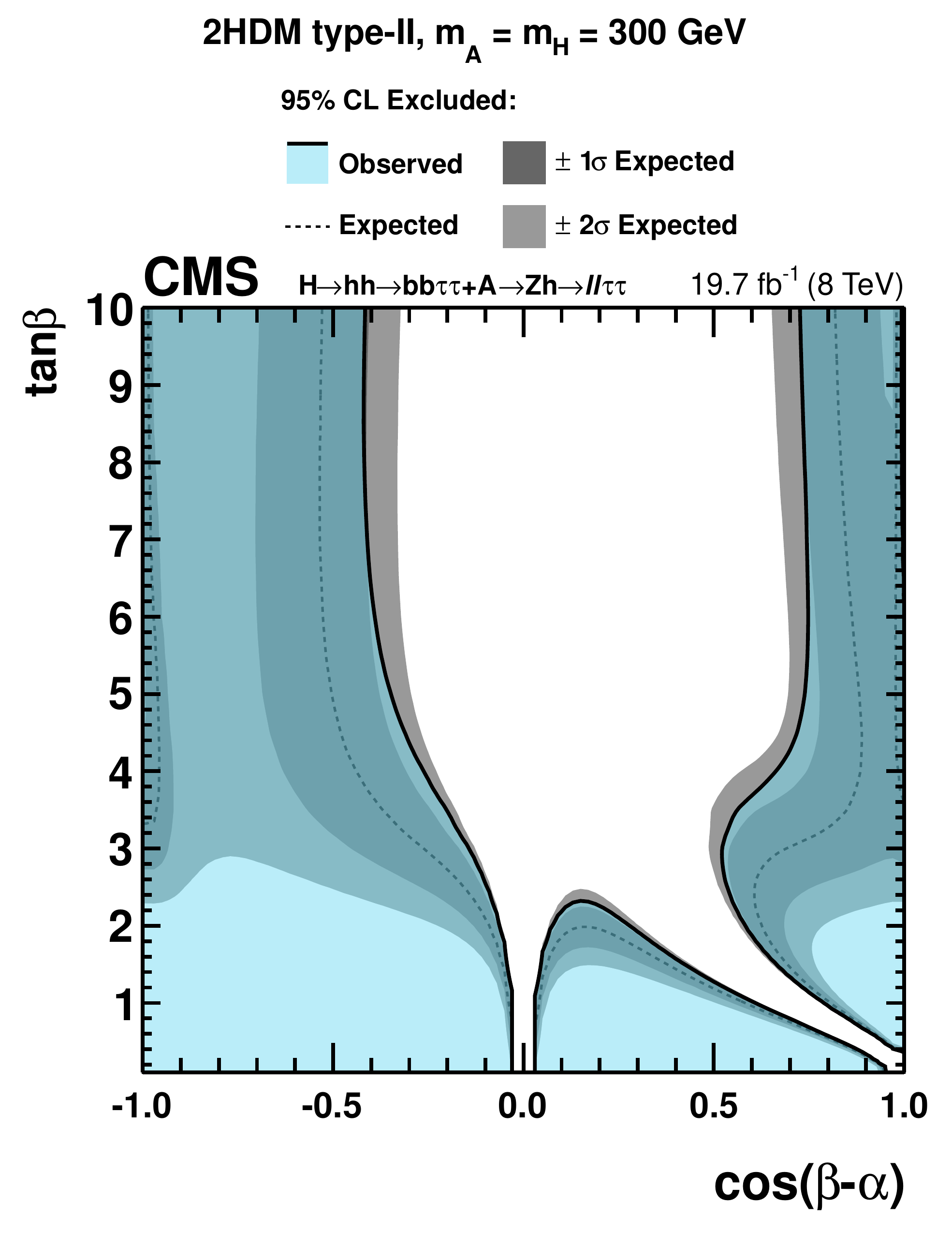}
\caption{The 95\% CL exclusion regions in the $\cos(\beta-\alpha)$ vs. $\tan\beta$ plane of 2HDM Type II
         model for $m_{\PA}=m_{\PH}=300\GeV$, combining the results of the $\PH\to\Ph\Ph\to\cPqb\cPqb\Pgt\Pgt$
         and $\PA\to\PZ\Ph\to\ell\ell\Pgt\Pgt$ analysis. The areas highlighted in
         blue bounded by the black curves mark the observed exclusion. The dashed curves and the grey bands show the expected exclusion limit with the relative uncertainty.
}
\label{fig:results_2HDMcmb}
\end{figure}

The cross-sections and branching fractions in the 2HDM were
calculated as described by the LHC Higgs Cross Section Working Group~\cite{Harlander:2013qxa,Harlander:2012pb}.
The exclusion regions, calculated using the combination of the $\PH\to\Ph\Ph\to\cPqb\cPqb\Pgt\Pgt$
and $\PA\to\PZ\Ph\to{\ell\ell}\Pgt\Pgt$ analyses,  in the
$\cos(\beta-\alpha)$ vs. $\tan\beta$ plane for such a Type II 2HDM scenario with a heavy Higgs boson mass of 300\GeV
are shown in
Fig.~\ref{fig:results_2HDMcmb}. This
can be compared to Fig.~5 in Ref.~\cite{Khachatryan:2015yjb}.

\section{Summary}
\label{sec:summary}

A search for a heavy scalar Higgs boson ($\PH$) decaying into a pair of SM-like Higgs bosons ($\Ph\Ph$) and a search for a heavy neutral pseudoscalar Higgs boson ($\PA$) decaying into a $\PZ$ boson and a SM-like Higgs boson ($\Ph$), have been performed using events recorded by the CMS experiment at the LHC. The dataset corresponds to an integrated luminosity of 19.7\fbinv, recorded at 8\TeV centre-of-mass energy in 2012.
No evidence for a signal has been found and exclusion limits on the production
cross section times branching fraction for the processes
$\PH\to\Ph\Ph\to\cPqb\cPqb\Pgt\Pgt$ and $\PA\to\PZ\Ph\to LL\Pgt\Pgt$ are presented. The results are also interpreted in the context of the MSSM and 2HDM models.

\begin{acknowledgments}
We congratulate our colleagues in the CERN accelerator departments for the excellent performance of the LHC and thank the technical and administrative staffs at CERN and at other CMS institutes for their contributions to the success of the CMS effort. In addition, we gratefully acknowledge the computing centres and personnel of the Worldwide LHC Computing Grid for delivering so effectively the computing infrastructure essential to our analyses. Finally, we acknowledge the enduring support for the construction and operation of the LHC and the CMS detector provided by the following funding agencies: BMWFW and FWF (Austria); FNRS and FWO (Belgium); CNPq, CAPES, FAPERJ, and FAPESP (Brazil); MES (Bulgaria); CERN; CAS, MoST, and NSFC (China); COLCIENCIAS (Colombia); MSES and CSF (Croatia); RPF (Cyprus); MoER, ERC IUT and ERDF (Estonia); Academy of Finland, MEC, and HIP (Finland); CEA and CNRS/IN2P3 (France); BMBF, DFG, and HGF (Germany); GSRT (Greece); OTKA and NIH (Hungary); DAE and DST (India); IPM (Iran); SFI (Ireland); INFN (Italy); MSIP and NRF (Republic of Korea); LAS (Lithuania); MOE and UM (Malaysia); CINVESTAV, CONACYT, SEP, and UASLP-FAI (Mexico); MBIE (New Zealand); PAEC (Pakistan); MSHE and NSC (Poland); FCT (Portugal); JINR (Dubna); MON, RosAtom, RAS and RFBR (Russia); MESTD (Serbia); SEIDI and CPAN (Spain); Swiss Funding Agencies (Switzerland); MST (Taipei); ThEPCenter, IPST, STAR and NSTDA (Thailand); TUBITAK and TAEK (Turkey); NASU and SFFR (Ukraine); STFC (United Kingdom); DOE and NSF (USA).

Individuals have received support from the Marie-Curie programme and the European Research Council and EPLANET (European Union); the Leventis Foundation; the A. P. Sloan Foundation; the Alexander von Humboldt Foundation; the Belgian Federal Science Policy Office; the Fonds pour la Formation \`a la Recherche dans l'Industrie et dans l'Agriculture (FRIA-Belgium); the Agentschap voor Innovatie door Wetenschap en Technologie (IWT-Belgium); the Ministry of Education, Youth and Sports (MEYS) of the Czech Republic; the Council of Science and Industrial Research, India; the HOMING PLUS programme of the Foundation for Polish Science, cofinanced from European Union, Regional Development Fund; the OPUS programme of the National Science Center (Poland); the Compagnia di San Paolo (Torino); the Consorzio per la Fisica (Trieste); MIUR project 20108T4XTM (Italy); the Thalis and Aristeia programmes cofinanced by EU-ESF and the Greek NSRF; the National Priorities Research Program by Qatar National Research Fund; the Rachadapisek Sompot Fund for Postdoctoral Fellowship, Chulalongkorn University (Thailand); and the Welch Foundation, contract C-1845.
\end{acknowledgments}
\clearpage
\bibliography{auto_generated}
\ifthenelse{\boolean{cms@external}}{}{
\clearpage
\appendix
\numberwithin{table}{section}
\numberwithin{figure}{section}
\input{supplemental_material}
}
\cleardoublepage \section{The CMS Collaboration \label{app:collab}}\begin{sloppypar}\hyphenpenalty=5000\widowpenalty=500\clubpenalty=5000\input{HIG-14-034-authorlist.tex}\end{sloppypar}
\end{document}

%% file: supplemental_material.tex
\section{Kinematic Fit}
\label{sec:kinfit}

In the analysed event topology $\PH \to \Ph \Ph \to\cPqb\cPqb\Pgt\Pgt$, the collinear approximation for the decay products of the $\Pgt$~leptons is assumed.
This is well motivated, since the $\Pgt$~leptons are highly boosted as they originate from a relatively heavy object compared to their own mass, $m_{\Ph}/m_{\Pgt}=70$.
Further, it is assumed that the reconstruction of the directions of all final state objects $\eta_i$ and $\phi_i$ with $i \in \{ \PQb_1,\PQb_2,\tau^{\text{vis}}_{1},\tau^{\text{vis}}_{2}\}$
is accurate and the uncertainties can be neglected compared to the uncertainties on the energy reconstruction.

Both, the pair of \cPqb~jets and the pair of $\Pgt$~leptons need to fulfil an invariant mass constraint
\begin{equation}
 m(\tau_1, \tau_2)= m(\PQb_1,\PQb_2)=m_{\Ph}=125\GeV.
\end{equation}
These two hard constraints reduce the number of fit parameters to two, chosen to be $E_{\PQb_1}$ and $E_{\tau_1}$.

For the two measured \cPqb~jet energies, the $\chi^2$ terms can be formulated as
\begin{equation}
\chi^2_{\PQb_{1,2}}=\left(\frac{E_{\PQb_{1,2}}^{\text{fit}}-E_{\PQb_{1,2}}^{\text{meas}}}{\sigma_{\PQb_{1,2}}}\right)^{2},
\end{equation}
where $E_{\PQb_{1,2}}^{\text{fit}}$ are the fitted and $E_{\PQb_{1,2}}^{\text{meas}}$ are the reconstructed \cPqb~jet energy, and $\sigma_{\PQb_{1,2}}$ describe the \cPqb~jet energy resolution.

In the decay of the two $\Pgt$~leptons at least two neutrinos are involved.  Thus there exists no good
measurement of the original $\Pgt$~lepton energies, but only lower energy limits.
For this reason, the $\Pgt$~lepton energies are constrained from the balance of the fitted heavy Higgs boson transverse momentum
\begin{equation}
  \vec{p}_{\mathrm{T,H}}^{\text{fit}} = \vec{p}_{\mathrm{T,\PQb_1}}^{\text{fit}} + \vec{p}_{\mathrm{T,\PQb_2}}^{\text{fit}} + \vec{p}_{\mathrm{T},\tau_1}^{\text{fit}} + \vec{p}_{\mathrm{T},\tau_2}^{\text{fit}}
\end{equation}
and the reconstructed transversal recoil
\begin{equation}
  \vec{p}_{\mathrm{T,recoil}}^{\text{meas}} =
  -\vec{p}_{\mathrm{T,miss}}^{\text{meas}} - \vec{p}_{\mathrm{T,\PQb_1}}^{\text{meas}}-\vec{p}_{\mathrm{T,\PQb_2}}^{\text{meas}}-\vec{p}_{\mathrm{T},\tau_1^{\text{vis}}}^{\text{meas}}-\vec{p}_{\mathrm{T},\tau_2^{\text{vis}}}^{\text{meas}} = - \vec{p}_{\mathrm{T,H}}^{\text{meas}}.
\end{equation}
Herein, $\vec{p}_{\mathrm{T},\text{miss}}^{\text{meas}}$ denotes the reconstructed missing momentum in the transverse plane, which has been determined from \MET reconstruction algorithms, as described in \refreco.
Any nonzero residual vector $\vec{p}_{\mathrm{T},\text{recoil}}^{\text{res}}= \vec{p}_{\mathrm{T,H}}^{\text{fit}}+\vec{p}_{\mathrm{T},\text{recoil}}^{\text{meas}}$ contributes to a $\chi^2$ term as follows
\begin{equation}
\chi^2_{\text{recoil}}=\vec{p}_{\mathrm{T},\text{recoil}}^{\text{res},T} \cdot \mathrm{V}^{-1}_{\text{recoil}} \cdot \vec{p}_{\mathrm{T},\text{recoil}}^{\text{res}}\,,
\end{equation}
where $\mathrm{V}_{\text{recoil}}$ denotes the covariance matrix of the reconstructed recoil vector.

The overall $\chi^2$ function finally reads,
\begin{equation}
	\chi^2 = \chi^2_{\PQb_1} + \chi^2_{\PQb_2} + \chi^2_{\text{recoil}}.
\end{equation}
After minimisation of this function by varying $E_{\PQb_1}$ and $E_{\tau_1}$, a very accurate reconstruction of the heavy Higgs boson mass ($M_{\PH}^{\text{kinfit}}$) is achieved.

%% file: HIG-14-034-authorlist.tex
\textbf{Yerevan Physics Institute,  Yerevan,  Armenia}\\*[0pt]
V.~Khachatryan, A.M.~Sirunyan, A.~Tumasyan
\vskip\cmsinstskip
\textbf{Institut f\"{u}r Hochenergiephysik der OeAW,  Wien,  Austria}\\*[0pt]
W.~Adam, E.~Asilar, T.~Bergauer, J.~Brandstetter, E.~Brondolin, M.~Dragicevic, J.~Er\"{o}, M.~Flechl, M.~Friedl, R.~Fr\"{u}hwirth\cmsAuthorMark{1}, V.M.~Ghete, C.~Hartl, N.~H\"{o}rmann, J.~Hrubec, M.~Jeitler\cmsAuthorMark{1}, V.~Kn\"{u}nz, A.~K\"{o}nig, M.~Krammer\cmsAuthorMark{1}, I.~Kr\"{a}tschmer, D.~Liko, T.~Matsushita, I.~Mikulec, D.~Rabady\cmsAuthorMark{2}, B.~Rahbaran, H.~Rohringer, J.~Schieck\cmsAuthorMark{1}, R.~Sch\"{o}fbeck, J.~Strauss, W.~Treberer-Treberspurg, W.~Waltenberger, C.-E.~Wulz\cmsAuthorMark{1}
% ...
\vskip\cmsinstskip
% ...
1:~~Also at Vienna University of Technology, Vienna, Austria\\
2:~~Also at CERN, European Organization for Nuclear Research, Geneva, Switzerland\\
% ...
